\documentclass[aps, prl, twocolumn,superscriptaddress,amsmath,amssymb,floatfix]{revtex4-1}

\usepackage{graphicx}
\usepackage[usenames]{color} 
\usepackage{comment} 
\usepackage{bm}
\usepackage{amsmath}

\newcommand{\be}{\begin{equation}} 
\newcommand{\ee}{\end{equation}}
\newcommand{\bea}{\begin{eqnarray}} 
\newcommand{\eea}{\end{eqnarray}}

\begin{document}

\title{Critical drying of liquids}

\author{Robert Evans}  
\author{Maria C. Stewart}  
\affiliation{H. H. Wills Physics Laboratory, University of Bristol, Royal Fort, Bristol BS8 1TL, United Kingdom.}
\author{Nigel B. Wilding} 
\affiliation{Department of Physics, University of Bath, Bath BA2 7AY, United Kingdom.} 

\begin{abstract}

We report a detailed simulation and classical density functional
theory study of the drying transition in a realistic model fluid
at a smooth substrate. This transition (in which the contact angle
$\theta\to 180^\circ$) is shown to be critical for both short ranged
and long-ranged substrate-fluid interaction potentials. In the latter
case critical drying occurs at exactly zero attractive substrate
strength. This observation permits the accurate elucidation of the
character of the transition via a finite-size scaling analysis of the
density probability function.  We find that the critical exponent
$\nu_\parallel$ that controls the parallel correlation length,
i.e. the extent of vapor bubbles at the wall, is over twice as large
as predicted by mean field and renormalization group calculations.
We suggest a reason for the discrepancy. Our findings shed
new light on fluctuation phenomena in fluids near hydrophobic and
solvophobic interfaces.

\end{abstract}

\maketitle

With new types of nanostructured hydrophobic substrates and coatings
finding application in systems such as microfluidic devices,
self-cleaning surfaces and chemical separation processes, there is
considerable interdisciplinary interest in the behavior of fluids in
contact with weakly attractive surfaces
\cite{Simpson:2015le,Li:2007aa,Ueda:2013aa,Checco:2014aa}. Thermodynamically,
the state of a liquid drop near a solid substrate (or `wall') is
characterized by the contact angle $\theta$ that the drop makes with
the surface. The weaker the wall-fluid attraction, the larger $\theta$
becomes. In the limit $\theta\to 180^\circ$ a fluid at vapor-liquid
coexistence undergoes a surface phase transition known as {\em drying}
whereby a macroscopic film of vapor ($v$) intrudes between the wall
($w$) and the bulk liquid ($l$); this is the analogue of the well
known wetting transition that occurs for strongly attractive surfaces
as $\theta\to 0$. Wetting has been studied in detail; see
\cite{Bonn:2009if} for a review and \cite{Friedman:2013nx} for a
recent investigation of water. Theory and simulation has often focused
on Ising models e.g. \cite{Binder:1989aa,Albano:2012db,Bryk:2013pi},
whose special symmetry implies that wetting and drying are equivalent.
However in real fluids, wetting and drying are distinct phenomena and
very little is known concerning the fundamental properties of either
transition. Previous work has led to long standing controversies in
particular as to whether the drying transition in model fluids is
first order or continuous (critical)
\cite{Swol:1989by,Henderson:1990nq,Swol:1991fq,Nijmeijer:1992fk,Nijmeijer:1991sw,Henderson:1992kk,Bruin:1995ud},
or even whether it exists at all
\cite{Ancilotto:2001uq,Oleinikova:2005uo}.  Accordingly there is a
need for clear elucidation of the nature of the approach to drying in
fluids, not just in thermodynamic terms, but also with regard to the
local density fluctuations that characterize the transition.

  The main barriers to computational progress in tackling drying in
  realistic fluids has been the dearth of techniques for
  locating surface phase transitions accurately, combined with the lack
  of rigorous measures for quantifying their key characteristics. In
  this Letter we deploy state-of-the-art Monte Carlo simulation
  techniques and classical density functional theory (DFT) together with
  a rigorously defined measure of the local compressibility to study a
  realistic model fluid near an attractive structureless wall. We begin
  by settling the long standing controversy concerning the order of the
  drying transition: For the (truncated) Lennard-Jones (LJ) fluid that
  we consider, drying is continuous (critical).  This is true for both a
  short-ranged (SR) and a long-ranged (LR) van der Waals wall-fluid
  interaction potential --a finding that contrasts with wetting in the
  same system which is a discontinuous transition for the LR wall-fluid
  potential but continuous for the SR case. Moreover, we show that for
  LR wall-fluid potentials, drying occurs at {\em zero} attractive wall
  strength. This represents the first instance of a surface phase
  transition in 3d whose parameters are exactly known and thus provides
  an opportunity to study a surface critical point free from uncertainty
  regarding its location (a problem that has previously plagued Ising
  model studies of critical wetting \cite{Bryk:2013pi}). By performing a
  finite size scaling (FSS) analysis of the density fluctuations that
  characterize the near critical region in the LR case, we demonstrate
  that critical drying in simulations is associated with a {\em single}
  divergent correlation length $\xi_\parallel$, that for density
  correlations parallel to the wall. The interfacial roughness
  $\xi_\perp$, arising from capillary wave fluctuations, is heavily
  dampened by finite-size effects to of order the particle diameter and
  plays no role in the FSS.  Our analysis allows us to estimate the
  effective critical exponent $\nu_{\parallel}$ describing the growth of
  $\xi_\parallel$.  We note that our 3d system is at the upper critical
  dimension and, in contrast to the case of SR wall-fluid interactions,
  a renormalization group (RG) analysis indicates \cite{Evans:SM2016} that
  the critical exponents should take their mean-field values. However,
  our simulation estimate of $\nu_{\parallel}$ is much larger than that
  predicted by mean-field and furthermore appears to be temperature
  dependent.

  The model we consider is a LJ fluid in a slit pore
  composed of a pair of structureless parallel walls of area $L^2$
  separated by a distance $D$; periodic boundary conditions apply in the
  directions parallel to the walls. Fluid-fluid interactions are
  truncated at $r_c=2.5\sigma$, where $\sigma$ is the LJ diameter, and particles interact with each wall via
  a wall-fluid potential $W(z)$, with $z$ the perpendicular
  particle-wall distance. We consider two forms for $W(z)$ commonly
  encountered in the adsorption literature: (i) the SR case of a hard
  wall plus square well potential of range $0.5\sigma$, and, (ii) the LR
  case of a hard wall plus a non-truncated long-ranged attraction
  decaying as $z^{-3}$. Both wall-fluid potentials are parameterized in
  terms of the well depth $\epsilon$.  To study these systems we deploy
  Grand Canonical Monte Carlo (GCMC) simulation and classical DFT. The
  latter approximates the repulsive LJ core as a hard core, whose free
  energy is treated via fundamental measure theory, while the attractive
  part of the LJ potential is treated in a mean field fashion
  \cite{Evans:1992jo,Evans:2015aa,Evans:SM2016}. The GCMC simulations impose the temperature $T$,
  chemical potential $\mu$ and the depth $\epsilon$ of $W(z)$, which we quote in units of $k_BT$. Flat
  histogram techniques \cite{berg1992} were used to record the
  local number density profile $\rho(z)$ and the overall
  number density $\rho$. All results were accumulated at liquid-vapor
  coexistence for two subcritical temperatures $T=0.775T_c$ and
  $T=0.842T_c$, with $T_c$ the bulk critical temperature known from
  previous work \cite{Wilding1995}. The coexistence value of $\mu$ was
  determined to high precision for a large fully periodic system using recently developed bespoke techniques
  \cite{Wilding:2016qr} which ameliorate the sampling problems at low $T$ and large volumes that arise from
  `droplet' transitions \cite{MacDowell:2004wj}.

\begin{figure}[h]
  \includegraphics[type=pdf,ext=.pdf,read=.pdf,width=0.94\columnwidth,clip=true]{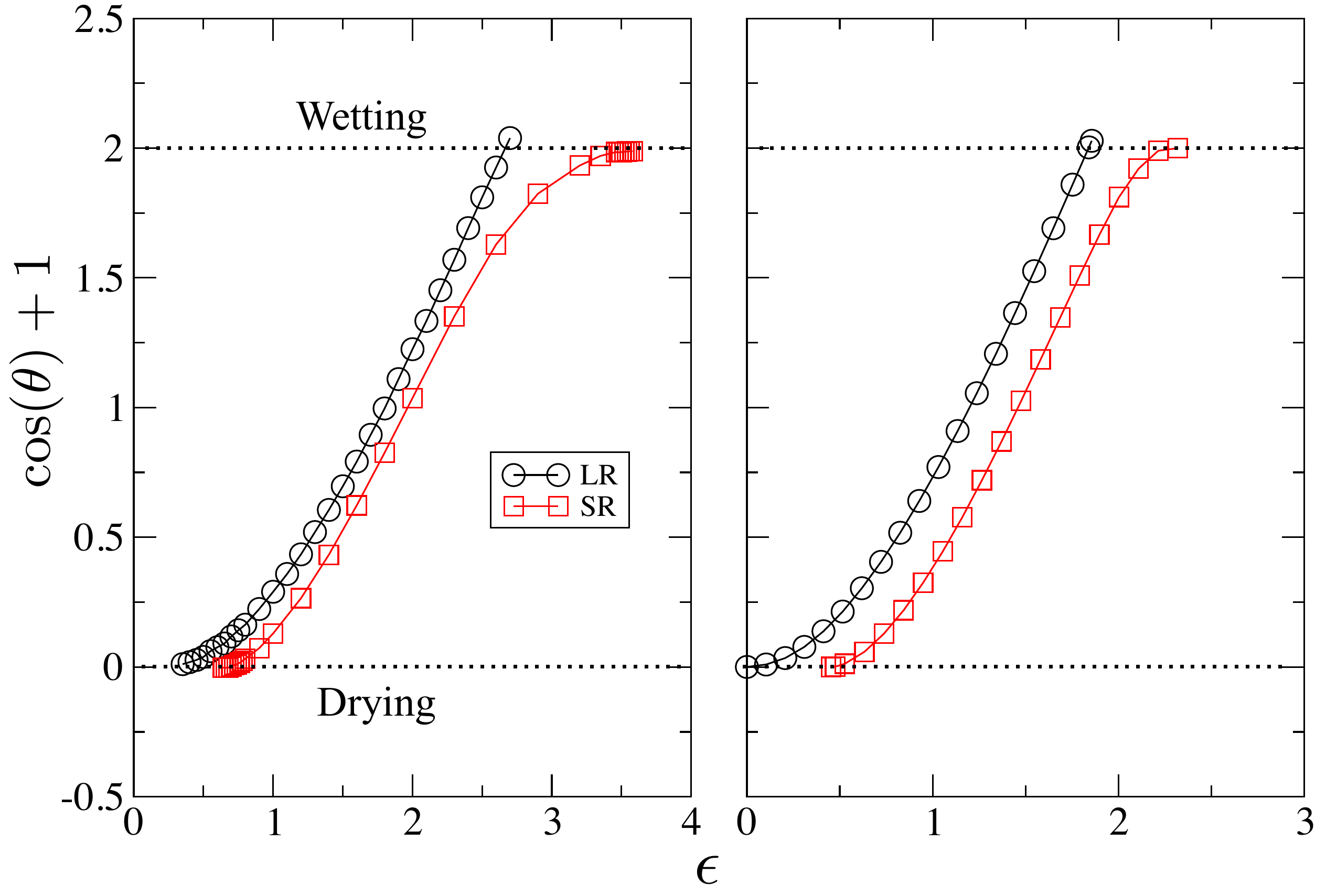}
  \caption{(Color online) Left: GCMC results for $\cos(\theta)+1$ versus
    $\epsilon$ for the SR and LR wall potential at $T=0.775T_c$, for
    $L=15\sigma, D=30\sigma$. For the LR case a FSS analysis
    of the GCMC data yields critical drying at $\epsilon_c=0$, as predicted by theory while for the SR case a FSS analysis gives critical drying at $\epsilon_c=0.52(2)$ and critical wetting at $\epsilon_c=4.25(5)$. Right: Corresponding DFT results. }
  \label{fig:costhetacompare}
  \end{figure}

  The dependence of the contact angle on the wall-fluid well depth
  $\epsilon$ was estimated for both the SR and LR wall-fluid potentials
  via direct measurements of the interfacial tensions appearing in
  Young's equation, $\gamma_{lv}\cos(\theta)=\gamma_{wv}-\gamma_{wl}$,
  using a method detailed elsewhere \cite{Muller:2000fv,
    Evans:2015wo}. The results are shown in
  Fig.~\ref{fig:costhetacompare} and span the range from wetting
  ($\cos(\theta)=1$) to drying ($\cos(\theta)=-1$). Interestingly the two
  forms of wall-fluid potential show distinct behavior.  For the SR
  case, both wetting and drying are continuous for this range of $W(z)$: $\cos(\theta)$ approaches the respective limits
  tangentially \cite{Dietrich:1988et}. For the LR case, the same is true for drying, but
  wetting is first order: $\cos(\theta)$ approaches unity with a
  non-zero linear slope. The DFT results in Fig.~\ref{fig:costhetacompare} display the same transitions as in simulation.

  In what follows we focus on drying in the LR case which is the
  situation most commonly studied in simulations of LJ fluids
  \cite{Fan:1993aa,Bryk:1999aa,Rane:2011ly} and of models of water
  \cite{Kumar:2013aa,Kumar:2013kx,Willard:2014aa}. From
  Fig.~\ref{fig:costhetacompare} it appears at first sight that here
  drying occurs at a non-zero (albeit small) value of
  $\epsilon$. However, while measurements of
  $\cos(\theta)$ are reliable indicators of the order of the
  transitions, they fail to provide accurate estimates for the critical
  well depth $\epsilon_c$. The problem goes well beyond that of the
  inherent difficulty of determining the point at which
  $\cos(\theta)=-1$ when the approach to this limit is
  tangential. Instead the main issue is one of critical finite-size
  effects which systematically shift the apparent critical point with
  respect to its true value. Accordingly a FSS analysis of the
  near-critical fluctuations is vital for determining accurately the
  drying point.

\begin{figure}[h]
  \includegraphics[type=pdf,ext=.pdf,read=.pdf,width=1.0\columnwidth,clip=true]{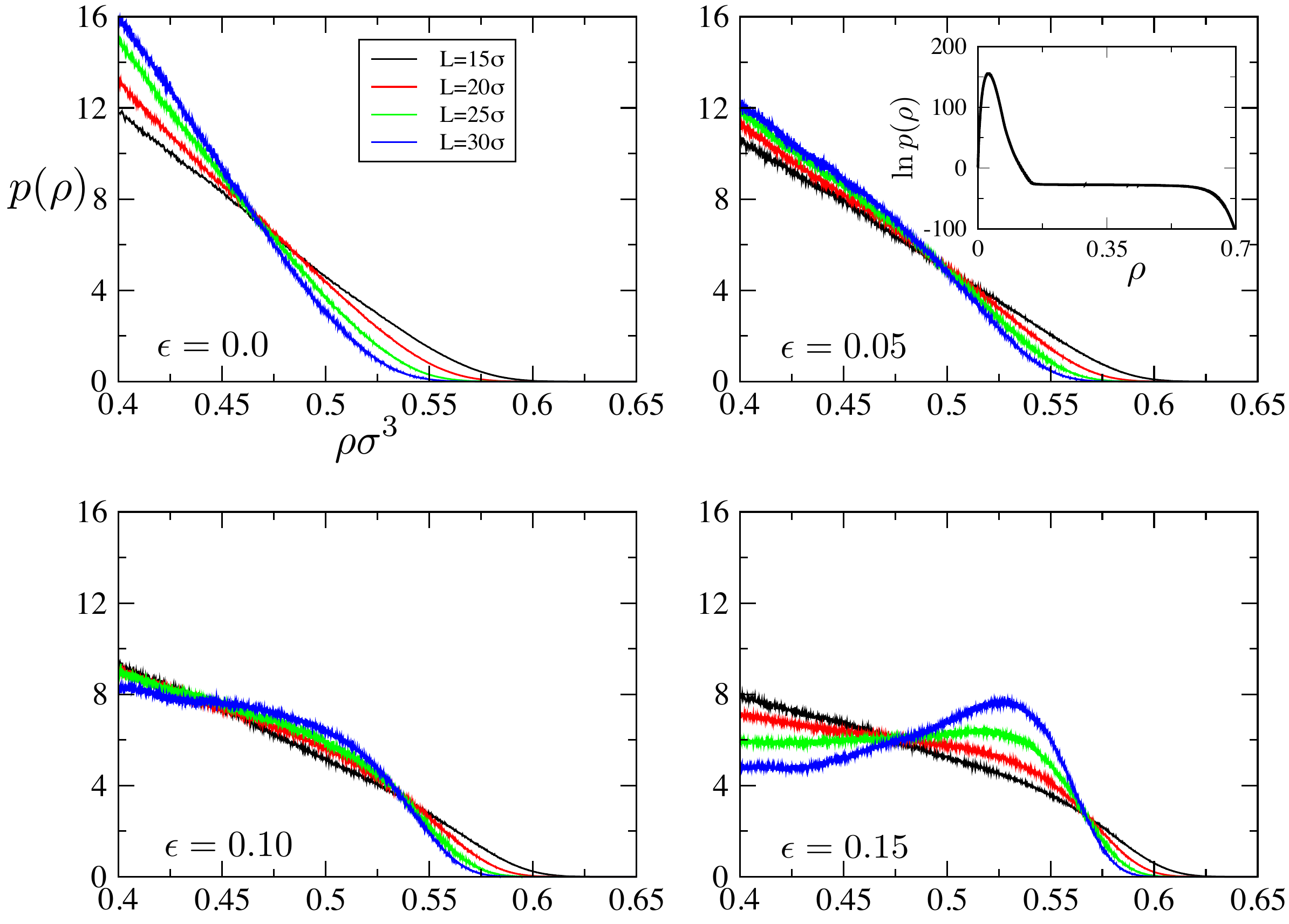}

  \caption{(Color online) GCMC results for $p(\rho)$ for the LR wall
    potential for $D=30\sigma$ and various $L$ at a selection of
    near-critical values of $\epsilon$.  Note that at small $\rho$ 
    capillary evaporation occurs, manifest as a gas-like peak in
    $p(\rho)$ as shown in the inset for $\epsilon=0.05, L=15\sigma$. }
  \label{fig:LRFS}
  \end{figure}

  Our approach is to examine the probability distribution function of
  the number density $p(\rho)$, and specifically its dependence on
  $\epsilon$ and the wall dimension $L$. Results are shown in
  Fig.~\ref{fig:LRFS} and reveal that for sufficiently large $\epsilon$
  and $L$, $p(\rho)$ exhibits a peak at high density. In the absence of
  finite-size effects, this peak corresponds to the liquid phase in
  contact with the wall and is a signature of partial drying
  ie. $\theta<180^\circ$. However, the situation is more subtle. On
  decreasing $\epsilon$, the peak in $p(\rho)$ disappears into a
  plateau. On further reducing $\epsilon$, $p(\rho)$ becomes
  monotonically decreasing with a bulge which gradually diminishes
  until, at $\epsilon=0$, the distribution comprises a linear part and a
  tail. The range of values of $\epsilon$ over which this scenario plays
  out decreases with increasing $L$. Only for $\epsilon=0$ is the form
  of $p(\rho)$ scale invariant, ie. no peak begins to form as $L$ is
  increased. Consequently this wall strength marks the critical drying
  point. Significantly, both DFT (c.f. fig.~\ref{fig:costhetacompare})
  and binding potential calculations \cite{Evans:SM2016} also predict
  critical drying for $\epsilon=0$.  When $\epsilon=0$, $W(z)$ reduces
  to the hard wall potential and complete drying occurs for all
  $T<T_c$. What is remarkable is that the transition is critical and
  occurs precisely at $\epsilon=0$ for all $W(z)$ exhibiting power-law
  decay \cite{Evans:SM2016}.

\begin{figure}[h]
  \includegraphics[type=pdf,ext=.pdf,read=.pdf,width=1.0\columnwidth,clip=true]{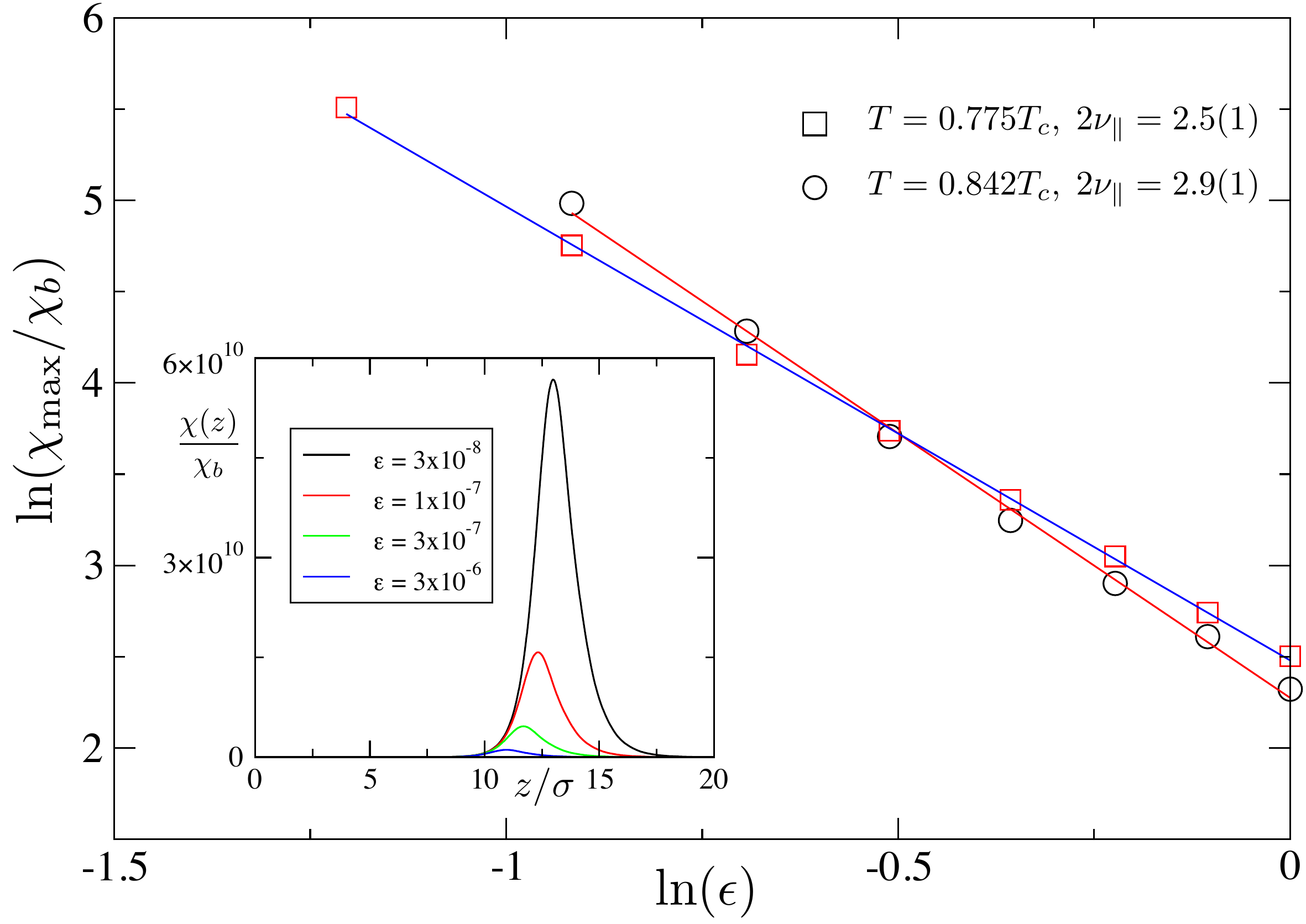}
  \caption{(Color online) GCMC measurements of the scaling of the peak in
    $\chi(z)/\chi_b$ with wall-fluid potential well depth $\epsilon$ for
    the LR wall-fluid potential. $\chi_b$ is the bulk liquid phase
    compressibility. The system size is $L=50\sigma, D=30\sigma$. Inset:
    DFT results for $\chi(z)/\chi_b$ for a single wall, showing the
    divergence as $\epsilon\to 0$. This occurs in the way binding
    potential arguments predict, i.e. $\ln(\chi(l)) \sim l$ with $l$ the
    drying layer thickness and $\chi(l) \sim \xi_\parallel^2$, see fig.~S1. of 
    \cite{Evans:SM2016}.}

  \label{fig:comp_peak_height}
  \end{figure}

  The fact that for a LR wall potential drying is critical
  with $\epsilon_c=0$ is confirmed by measurements of the
  compressibility profile $\chi(z)\equiv\partial \rho(z)/\partial \mu$.
  This quantity was introduced previously
  \cite{Tarazona:1982aa,Evans:1990aa} and has subsequently proven a sensitive measure of
  the link between the contact angle $\theta$ and the local
  structure near hydrophobic or solvophobic surfaces
  \cite{Evans:2015aa,Evans:2015wo}. Its form probes the transverse
  density-density correlation function and thus the correlation length
  $\xi_\parallel$.  GCMC measurements of the maximum of $\chi(z)$ are
  shown in Fig.~\ref{fig:comp_peak_height} and demonstrate a power law
  divergence as $\epsilon$ is reduced to zero, implying that
  $\xi_\parallel$ diverges at this wall strength. This divergence is
  confirmed by DFT measurements of $\chi(z)$ as shown in the inset.

  The form of $p(\rho)$ at $\epsilon=0$ (Fig.~\ref{fig:LRFS}),
  corresponding to a hard wall, represents a hallmark of critical drying
  and yields fundamental insight concerning its character. It comprises
  a linearly sloped part at lower density plus a tail at higher
  densities. With increasing $L$, the tail density shifts to lower
  values. Interestingly, the form and $L$-dependence of
  $p(\rho|\epsilon_c)$ cannot be rationalized in terms of a FSS ansatz
  previously proposed for critical wetting in $3d$ Ising models
  \cite{Albano:2012db,Bryk:2013pi}. That theory presumes the critical
  divergence of not just $\xi_\parallel$, but also of the perpendicular
  correlation length $\xi_\perp$ which measures the roughness of the
  emerging liquid-vapor interface due to capillary fluctuations. While
  our measurements of $\chi(z)$ provide ample evidence for a divergent
  $\xi_\parallel$, we find no signs that $\xi_\perp$ is large in our
  simulations. This is because of the extremely strong finite-size
  dampening of the surface roughness for $d=3$. General capillary wave
  arguments e.g. \cite{Gelfand:1990ve,Dietrich:1988et,Evans:1992jo} for
  a single unbinding vapor-liquid interface predict that
  $\xi_\perp\simeq \sqrt{(k_BT/2\pi\gamma_{lv})\ln (L/\xi_b)}$. Thus the
  interfacial roughness depends on the finite {\em lateral} dimension of
  the system.  Given the strength of this dampening, one cannot expect
  $\xi_\perp$ to become large on the scale of the particle diameter (or
  indeed the bulk correlation length $\xi_b$) for currently accessible
  simulation sizes.

\begin{figure}[h]
  \includegraphics[type=pdf,ext=.pdf,read=.pdf,width=1.0\columnwidth,clip=true]{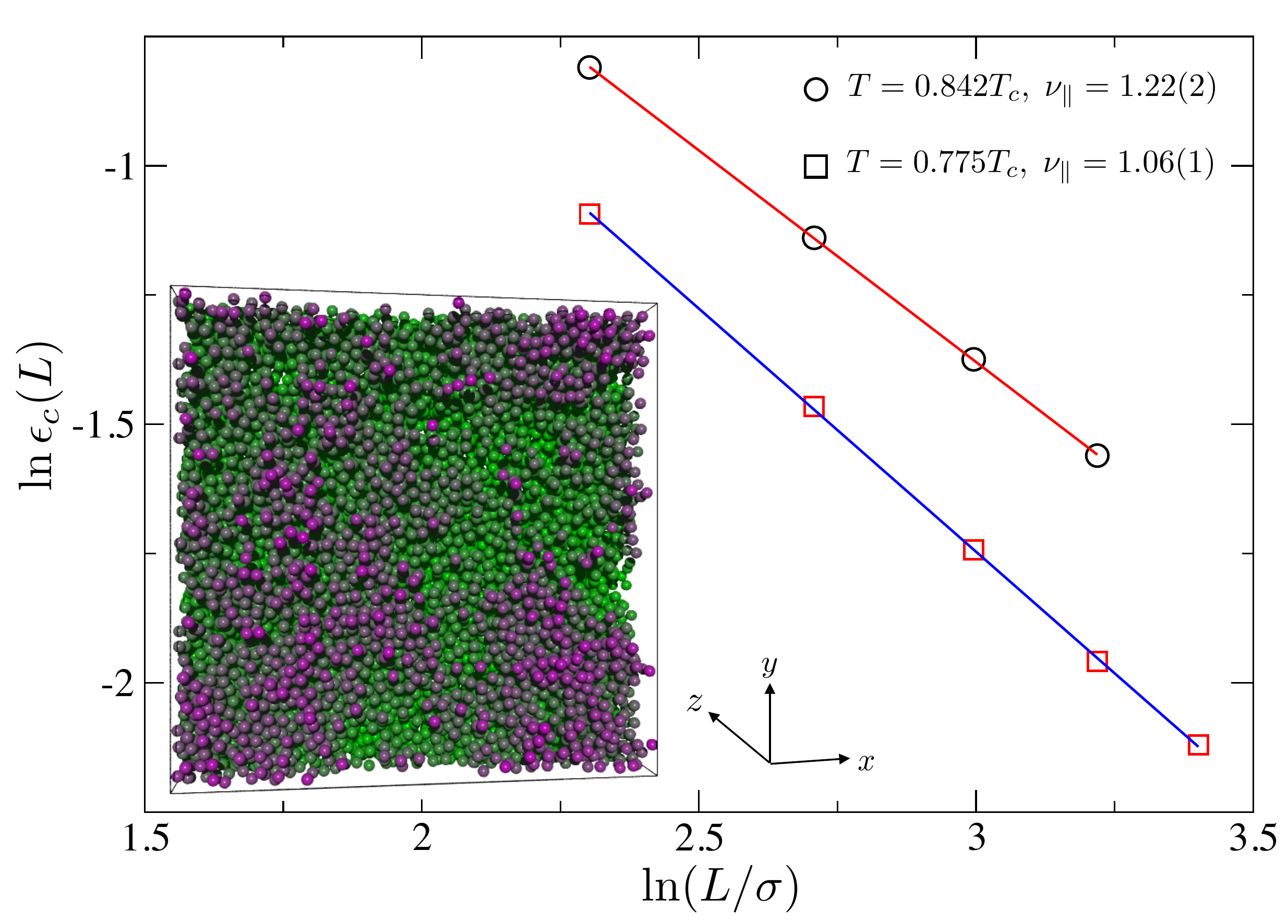}
  \caption{The scaling of $\epsilon_c(L)$, i.e. the wall strength at
    which a peak appears in $p(\rho)$, as a function of $L$ for the LR
    wall-fluid potential. Data are shown for two subcritical
    temperatures. Inset: Simulation snapshot for a system with 
    $L=40\sigma, \epsilon=0.2$. Particles are color coded
    according to their distance from the wall at $z=0$. A large
    correlation length is manifest in the vapor close to the wall; see
    SM \cite{Evans:SM2016}.}
  \label{fig:lpeaks}
  \end{figure}

  These observations, together with the results of Figs.~\ref{fig:LRFS}
  and \ref{fig:comp_peak_height}, imply the following picture for
  critical drying in simulations of 3d systems. As $\epsilon\to
  \epsilon_c^+$, bubbles of vapor form at the wall whose lateral size
  corresponds to $\xi_\parallel\sim
  (\epsilon-\epsilon_c)^{-\nu_{\parallel}}$ (cf. the snapshot in 
  fig.~\ref{fig:lpeaks} and the movie in the SM\cite{Evans:SM2016}), but whose perpendicular lengthscale
  remains microscopic. As $\xi_\parallel$ approaches $L$, the liquid
  unbinds from the wall to form a `slab', surrounded by
  vapor. Essentially this process can be viewed as premature drying
  induced by the finite system size. The slab surface is rather sharp and
  localized due to the dampening of interfacial roughness and the slab
  thickness (in the $z$-direction) is therefore proportional to
  $\rho$. Accordingly, the linear decrease of $p(\rho|\epsilon=0)$ seen
  at low to moderate densities in fig.~\ref{fig:LRFS} arises simply from
  the `entropic repulsion' of the slab and the wall:
  the number of positions for the slab center along the $z$ axis that
  are allowed by the presence of the wall, varies linearly with slab
  thickness. The high density tail of $p(\rho)$ on the other hand
  reflects the free energy cost of pushing the liquid up against the
  wall, the act of which quenches the parallel density
  fluctuations.  Its $L$ dependence arises --as shown in the
  SM \cite{Evans:SM2016}-- from a constant
  repulsive pressure on the liquid-vapor interface by the wall, 
  giving rise to a force which scales simply with the wall area $L^2$.

  Neither the fluctuation in the thickness of the unbound
  liquid slab occurring at low-moderate densities, nor the high
  density tail is directly associated with criticality, and thus one
  cannot expect $p(\rho)$ to exhibit non-trivial FSS behavior as a
  whole.  Rather, the signature of near critical fluctuations is
  manifest in the density range where the liquid is still (weakly) bound
  to the wall but exhibits strong parallel density fluctuations.  This
  correspond to the liquid peak in Fig.~\ref{fig:LRFS}, the height of
  which depends on $\xi_\parallel$ and vanishes when
  $\xi_\parallel\approx L$ allowing the liquid slab to unbind from the
  wall. Simple FSS dictates that this vanishing occurs not at $\epsilon_c$
  but at the larger effective value $\epsilon_c (L) = \epsilon_c +
  aL^{-1/\nu_\parallel}$ (which corresponds also to the wall strength at
  which the surface tension measurements with Young's equation, predict
  $\theta=180^\circ$). The critical wall strength $\epsilon_c$ can
  differ substantially from $\epsilon_c(L)$ and is determined most
  accurately as the largest value of $\epsilon$ for which $p(\rho)$
  assumes an $L$-independent form. However, in contrast to the rich
  structure of the density distribution at bulk criticality
 \cite{Wilding1995} the novel
  feature of critical drying is the surprising simplicity of
  $p(\rho|\epsilon_c)$.


  We have determined the value of $\nu_\parallel$ via the
  anticipated FSS $\epsilon(L)\sim L^{-1/\nu_\parallel}$;
  $\epsilon_c=0$ for the LR case.  For a number of choices of $L$ we
  measured $\epsilon(L)$ accurately (using histogram extrapolation
  techniques) from the vanishing of the liquid peak of $p(\rho)$
  (cf. fig.~\ref{fig:LRFS}). As fig.~\ref{fig:lpeaks} shows, we do
  indeed see power law scaling, from which we can extract an estimate
  of $\nu_\parallel$. Interestingly, however, this estimate exceeds
  the prediction $\nu_\parallel=0.5$ of mean field and RG
 theories (see SM \cite{Evans:SM2016}) by over a factor of two
  and additionally appears to show a clear temperature
  dependence. This discrepancy with theory is further mirrored in the
  behavior of $\chi(z)$ (fig.~\ref{fig:comp_peak_height}) for which
  one expects \cite{Evans:SM2016} that $\chi_{\rm
    max}\sim(\epsilon-\epsilon_c)^{-2\nu_\parallel}$.  Here too the
  simulation estimates of $\nu_\parallel$ are over twice the
  theoretical prediction and show a clear temperature dependence.

  We summarize and discuss our findings.  A realistic model liquid in
  contact with a substrate that exerts a long-ranged van der Waals
  attraction undergoes a critical drying transition at zero attractive
  wall strength. From the general theory \cite{Evans:SM2016}, we can
  infer that the same transition, at $\epsilon=0$, should occur for
  models of water at such walls. Indeed the occurrence of critical
  drying would account for recent results
  \cite{Kumar:2013aa,Kumar:2013kx,Willard:2014aa,Godawat:aa} displaying
  very large contact angles and enhanced fluctuations in simulations of
  water at strongly hydrophobic LR substrates.

  Analysis of density fluctuations provides fresh insight
  into the nature of critical drying, revealing that (in simulations
  at least) there is only one divergent correlation length,
  $\xi_\parallel$, associated with the growth of vapor bubbles at the
  wall.  Of course capillary wave theory predicts that $\xi_\perp$
  diverges for a free interface in the absence of gravity, or at an
  infinite single wall in the limit of wetting/drying, but it seems
  one cannot observe a macroscopically large $\xi_\perp$ in
  fluid simulations, which therefore miss a key element of the
  theoretical picture \cite{thickfoot}.  It is tempting to speculate that a single
  diverging $\xi_\parallel$ could imply that critical drying in
  simulations is effectively controlled by the 2D Ising fixed point
  for which $\nu=1$. This value is indeed much closer to our estimate
  of $\nu_\parallel$ than the predictions of RG theory for the LR
  case.  Clearly further work is required to address these subtle but
  important issues.

  Our methods for locating and characterizing critical drying
  should prove useful for elucidating critical
  wetting transitions in $d=3$. Here fundamental questions remain
  regarding the relationship between simulation results and theoretical
  predictions
  \cite{Albano:2012db,Bryk:2013pi,Parry:2008ef,Parry:2008rp,Parry:2009aa}.
  In fig.~\ref{fig:costhetacompare} our results for $\cos(\theta)$ for a
  SR (square-well) wall indicate critical wetting. Preliminary
  investigations \cite{Evans:ab} of this system reveal closely analogous
  phenomenology to that seen at drying, namely a gas-peak in $p(\rho)$
  which gradually disappears on increasing $\epsilon$ until, at the
  wetting point, $p(\rho)$ assumes a scale invariant form comprising a
  low density tail and a linear part extending to high density. The
  implication is that like critical drying, critical wetting in
  simulations will occur in the absence of a large $\xi_\perp$.

  Our findings settle the long standing controversy regarding the
  order of the drying transition
  \cite{Swol:1989by,Henderson:1990nq,Swol:1991fq,Nijmeijer:1992fk,Nijmeijer:1991sw,Henderson:1992kk,Bruin:1995ud}. Furthermore
  they help explain the original misconception. This arose, we believe,
  because for fluids in a slit pore the liquid phase is metastable
  with respect to capillary evaporation (cf. the gas peak in the
  inset of fig.~\ref{fig:LRFS}). As $\epsilon\to \epsilon_c(L)^+$, the
  liquid unbinds from the wall and the liquid-vapor interface wanders
  towards the slit center where it annihilates with its
  counterpart from the other wall to form a pure gas phase. In the
  absence of the insights provided by the present work, it is easy to
  mistake this discontinuous evaporation for the critical
  surface phase transition that precipitates it
  \cite{Swol:1989by,Henderson:1990nq,Swol:1991fq}. Note, however, that
  since the results of fig.~\ref{fig:LRFS} focus on the regime of
  moderate to large $\rho$ they are unaffected by evaporation
  \cite{Note1}.  

Finally, as regards the experimental relevance of our findings, the
observation that the drying transition in liquids is critical
irrespective of the range of the wall-fluid interactions, should prove
important when interpreting observations of the properties of fluids
near hydro- or solvo-phobic interfaces, in which there is growing
technological \cite{Simpson:2015le,Li:2007aa,Ueda:2013aa} and
fundamental \cite{Mezger:2011aa,Uysal:2013aa} interest. We do not
expect the basic phenomenology of critical drying to be altered if one
considers a substrate corrugated on the atomic scale rather than a
planar one. It remains to be seen to what extent the phenomenology
applies for nanostructured surfaces with larger characteristic
periods. Although real hydrophobic surfaces never quite attain contact
angles $\theta=180^\circ$, the effects of criticality should extend
over a wide range of $\theta<180^\circ$
\cite{Evans:2015wo,Evans:2015aa} and experiments such as those of
ref.~\cite{Nygaard:2016aa} might be able to confirm the existence of
enhanced density fluctuations in the vicinity of a hydrophobic
substrate.

  \acknowledgments
R.E. acknowledges Leverhulme Trust grant EM-2016-031.

\newpage

\makeatletter
\renewcommand{\fnum@figure}{\figurename~S\thefigure}
\makeatother
\setcounter{figure}{0}
\section{Supplementary Material}

\vspace*{-2mm}
\subsection{Simulation Details}
\vspace*{-2mm}

For our LJ fluid, particles interact via the potential, 
\be
\phi_{\rm att}(r)=\left \{ \begin{array}{ll}
 4\epsilon_{LJ}\left[\left(\frac{\sigma}{r}\right)^{12}-\left(\frac{\sigma}{r}\right)^{6}\right], & r\le r_c \, ,\\
0, & r>r_c,\\
\end{array}
\right.
\label{eq:Simpot}
\ee
with $\epsilon_{LJ}$ the well-depth of the potential and $\sigma$ the
LJ diameter. We choose $r_c=2.5\sigma$, for which criticality occurs
\cite{Wilding1995} at $k_BT_c=1.1876(3) \epsilon_{LJ}$. We work at
$k_BT=0.91954\epsilon_{LJ}=0.775T_c$ for which coexistence occurs at
$\beta\mu_{co}=-3.865950(20)$, with coexistence densities $\rho_l\sigma^3=0.704(1)$
and $\rho_v\sigma^3=0.0286(2)$; and also at $k_BT_c=1.0\epsilon_{LJ}=0.842T_c$ for
which $\beta\mu_{co}=-3.457131(25)$, $\rho_l\sigma^3=0.653(1),\rho_l\sigma^3=0.0504(3)$.

We employ two types of wall-fluid potential in our GCMC
simulations. The SR potential is a square-well given by

\be
W_{\rm SR}(z)=\left \{ \begin{array}{ll}
\infty, \mbox{\hspace{4mm}}   &  z\le 0  \\
 -\epsilon, & 0<z<\sigma/2 \, ,\\
0, & z>\sigma/2,\\
\end{array}
\right.
\label{eq:SRpot}
\ee
where $\epsilon$ is the well-depth. The LR potential is given by

\be
W_{\rm LR}(z)=\left \{ \begin{array}{ll}
\infty, \mbox{\hspace{4mm}}   &  z\le 0  \\
 \epsilon_w\epsilon_{LJ}\left[\frac{2}{15}\left(\frac{\sigma}{\tilde z}\right)^9-\left(\frac{\sigma}{\tilde z}\right)^3\right], & z>0 \, ,\\
\end{array}
\right.
\label{eq:LRpot}
\ee
where $\tilde z=z+(2/5)^{1/6}\sigma$, use of which shifts the minimum of the $9$-$3$ potential to the hard wall at $z=0$.
$\epsilon_w$ is a dimensionless measure of the strength of the wall-fluid
attraction. At the minimum of (\ref{eq:LRpot}) the value of the wall-fluid
potential is $-1.0541\epsilon_w\epsilon_{LJ}=-\epsilon$.

\vspace*{-2mm}
\subsection*{Binding potential analysis for the LR case}
\vspace*{-2mm}

We follow the standard treatment, e.g. [27], of wetting/drying
transitions and consider $\omega^{ex}(l)$, the excess grand potential
per unit surface area, as a function of the thickness $l$ of the
wetting/drying layer. For a truncated LJ model adsorbed at a single
wall exerting the potential (\ref{eq:LRpot}) we expect

\be
\omega^{ex}(l)=\gamma_{wv}+\gamma_{lv}+\omega_B(l)+\delta\mu (\rho_l-\rho_v)l
\label{eq:grandpot}
\ee
with the binding potential 

\be
\omega_B(l)=a\exp{(-l/\xi_b)}+bl^{-2}+{\rm H.O.T.}
\label{eq:bpot}
\ee
$\rho_l$ and $\rho_v$ are the liquid and vapor densities at
coexistence, $\delta\mu=\mu-\mu_{co}\ge 0$ is the deviation of the
chemical potential from its value at coexistence and we have
specialized now to the case of drying, i.e.  $l$ is the thickness of a
layer of vapor that can intrude between the weakly attractive wall and
the bulk liquid at $z= \infty$. In the limit of complete drying, at
$\delta\mu=0^+$, $l$ diverges and the $wl$ interface is a composite of
the $wv$ and $lv$ interfaces. In this limit
$\gamma_{wl}=\gamma_{wv}+\gamma_{lv}$, i.e. $\cos(\theta)= -1$. The
binding potential in (4) has two leading contributions. The
exponential term accounts for SR fluid-fluid interactions; $\xi_b$ is
the true correlation length of the bulk phase that wets, in our case
the vapor, and $a$ is a positive coefficient.  The term $bl^{-2}$ is
associated with the $z^{-3}$ decay of $W_{LR}(z)$ in (\ref{eq:LRpot}); it arises from
dispersion (van der Waals) forces between the substrate and the
fluid. The higher order terms in (\ref{eq:bpot}) include higher inverse powers such as $cl^{-3}$ 
as well as more rapidly decaying exponentials. We ignore these
in the subsequent analysis.  Making a straightforward sharp-kink
approximation, or Hamaker type calculation, e.g. [27], yields

\be
b=-(\rho_l-\rho_v)\epsilon_w\epsilon_{LJ}\sigma^3/2
\label{eq:b}
\ee
Since $b<0$ for all $T<T_c$, minimizing (\ref{eq:grandpot}) w.r.t. $l$ at $\delta\mu=0^+$, leads to a
finite value for the equilibrium thickness:

\be
\frac{-l_{eq}}{\xi_b}=\ln\epsilon_w-3\ln\left(\frac{l_{eq}}{\xi_b}\right)+{\rm constants};\hspace*{1mm} \delta\mu=0^+
\label{eq:leq}
\ee

A formula equivalent to (\ref{eq:leq}) was derived by Nightingale {\em et
    al.} (see Eq. 6 of \cite{Nightingale:1983ty}) in a
    study of critical wetting in systems with LR forces. Those authors
    considered only the case where $\epsilon_w >0$ and concluded there
    was no wetting, critical or first order. Here we focus on the
    situation where $\epsilon_w \to 0^+$, $b\to 0^-$ and $l_{eq}$ diverges continuously. Note
    that for $\epsilon_w =0$, $W_{LR}(z)$ reduces to the planar hard-wall potential
    and minimization of (\ref{eq:grandpot}) then yields $-l_{eq}/\xi_b=\ln(\delta\mu)+{\rm const}$, the mean-field (MF) result
    appropriate for complete drying from off-coexistence, for all $T<T_c$, e.g.~\cite{Evans:1992jo,Dietrich:1988et,Evans:1990aa}.  

Using (\ref{eq:grandpot},\ref{eq:bpot}) we can calculate several properties
    and examine these, within MF, in the approach to critical drying
    $\epsilon_w\to 0^+$. The local compressibility, evaluated for $z\approx l_{eq}$, is given
    by \cite{Evans:2015aa,Evans:1990aa}

\be
\chi(l_{eq})=\left(\frac{\partial\rho(z)}{\partial\mu}\right)_{z=l_{eq}}\sim-\rho^\prime(l_{eq})\left(\frac{\partial l_{eq}}{\partial \mu}\right)
\label{eq:Chi}
\ee
where the prime denotes differentiation w.r.t. $z$. From (\ref{eq:grandpot}) it follows that, at leading order, 

\be
\left(\frac{\partial l_{eq}}{\partial\mu}\right)=-\frac{\xi_b^2}{a}(\rho_l-\rho_v)\exp({l_{eq}/\xi_b}); \hspace*{1mm}\delta\mu=0^+
\label{eq:leqder}
\ee
Capillary wave arguments predict that in the limit of critical drying
$\rho^\prime(l_{eq})\sim\xi_\perp^{-1}$, where $\xi_\perp$ is the
interfacial roughness. Within MF, $\xi_\perp^{-1}$ is non-zero, and using
(\ref{eq:leq}) we deduce 

\be
\ln\chi(l_{eq})\sim \frac{l_{eq}}{\xi_b} + {\rm const.};\hspace*{2mm} \delta\mu=0^+
\label{eq:logchi}
\ee
The predictions (\ref{eq:leq}) and (\ref{eq:logchi}) were tested using the microscopic DFT, as described below.

We can also extract the correlation length $\xi_\parallel$ that
describes density-density correlations parallel to the wall. General
arguments, see Refs.~\cite{Evans:2015aa,Evans:1990aa} and references therein, predict that
$\xi_\parallel^2$ diverges in the same way as the surface excess
compressibility defined as

\bea
\chi_{ex}&\equiv & \left(\frac{\partial\Gamma}{\partial\mu}\right)=\frac{\partial}{\partial\mu}\int_0^\infty dz(\rho(z)-\rho_b)\nonumber\\
    &=&\int_0^\infty dz(\chi(z)-\chi_b)
\eea
where, $\rho_b=\rho(\infty)$ is the bulk density and $\Gamma$ is the
Gibbs adsorption. Since $\chi_{ex}$ is proportional to $-\partial l_{eq}/\partial\mu$ it follows
from (\ref{eq:leqder},\ref{eq:leq}) that $\xi_\parallel$ diverges as

\be
\xi_\parallel\sim\epsilon_w^{-1/2}(-\ln\epsilon_w)^{3/2}; \hspace*{1mm}\delta\mu=0^+
\label{eq:xipar}
\ee 
in the limit $\epsilon_w \to 0^+$. The same result is obtained
from standard binding potential considerations [27] where one has
$\xi_\parallel^{-2}\propto \partial^2\omega_B(l)/\partial l^2$ at
$l=l_{eq}$.

The variation of $\cos(\theta)$ close to critical
drying is determined by $\omega_B(l_{eq})$ at $\delta\mu=0^+$, i.e. the
singular part of the surface excess free energy $\gamma^{sing}$. Using Young’s
equation one finds $1+ \cos(\theta)=-\omega_B(l_{eq})/\gamma^{lv}$ and for the present binding potential (\ref{eq:bpot})
we obtain

\be
1+\cos(\theta)\sim\epsilon_w(-\ln\epsilon_w)^{-2}
\label{eq:costheta}
\ee
in the limit $\epsilon_w\to 0^+$. This result is striking. Were the
logarithm not present in (\ref{eq:costheta}) the theory would predict $1+ \cos(\theta)$
vanishing linearly with $\epsilon_w$, a signature of a 1st order drying
transition. It is only the presence of the logarithm that ensures a
continuous (critical) transition. The critical exponent $\alpha_s$,
defined by the vanishing of the singular part of the surface excess
free energy $\gamma^{sing}\sim \epsilon_w^{2-\alpha_s}$, clearly takes the value $\alpha_s =1$, with log
corrections, in this particular case. The situation is similar to that
for complete drying from off-coexistence where for a planar hard-wall,
say, $\gamma^{sing}\sim\delta\mu\ln\delta\mu$, $\delta\mu\to 0^+$. 

It is important to distinguish the MF scenario presented above from
that corresponding to a SR wall-fluid potential such as (1). In the SR
case it is well-known, e.g.~[27,41], that the second inverse power-law
term in (\ref{eq:bpot}) must be replaced by a H.O. term proportional
to $\exp(-2l/\xi_b)$ while the coefficient of the leading
$\exp{(-l/\xi_b)}$ term now depends on $\epsilon_w$:
$a(\epsilon_w)\sim (\epsilon_w- \epsilon^{MF}_{wc})$ where
$\epsilon^{MF}_{wc}$ is the strength of the wall-fluid attraction at
which critical drying occurs in MF. Defining
$\delta\epsilon_w=\epsilon_w- \epsilon^{MF}_{wc}$, MF analysis for the SR
case yields, for $\delta\mu=0^+$,

\be
\frac{-l_{eq}}{\xi_b}\sim\ln(\delta\epsilon_w); \hspace*{2mm} \chi(l_{eq})\sim(\delta\epsilon_w)^{-2}; \hspace*{2mm}\xi_\parallel\sim (\delta\epsilon_w)^{-1}
\ee

and

\be
1+\cos(\theta)\sim(\delta\epsilon_w)^2 \hspace*{2mm} {\rm or} \hspace*{2mm} \alpha_s=0;
\ee
These results are clearly very different from those we obtained above for the LR case.

\vspace*{-2mm}
\subsection*{A Renormalization Group (RG) treatment of fluctuations}
\vspace*{-2mm}

The analysis described above was strictly MF; this omits some of the effects of capillary wave (CW) fluctuations. For example, for infinite surface area, MF predicts a sharp interface with $\xi_\perp$ finite in all dimensions $d$ whereas, in reality, we expect $\xi_\perp$ to diverge for $d\le 3$. An important early attempt to incorporate CW fluctuations was that of Brezin et al. \cite{Brezin:1983dn} who introduced a RG treatment for the case of SR forces where the upper critical dimension is $d=3$ for both critical wetting and complete wetting from off-coexistence. We follow their methodology for our binding potential (\ref{eq:bpot}).

First we invoke the hyperscaling relation $(2-\alpha_s) =(d-1)\nu_\parallel$, where $\nu_\parallel$ is the critical exponent for $\xi_\parallel$, insert the MF exponents given above and deduce that the upper critical dimension is, once again, $d =3$. Next we introduce the standard, dimensionless parameter   $\omega=(4\pi\beta\gamma_{lv}\xi_b^2)^{-1}$, with $\beta=(k_BT)^{-1}$, that measures the strength of CW fluctuations. The RG treatment then implies we should consider an effective binding potential (renormalized) at the scale $\xi_\parallel$:
\be
\omega_{\xi_\parallel}(l)=a \xi_\parallel^\omega\exp{(-l/\xi_b)}+bl^{-2}+\delta\mu(\rho_l-\rho_v)l
\label{eq:omegaxi}
\ee
The exponential term is renormalized but the remaining power-law terms are not; in particular the coefficient $b$ is assumed to be unchanged. Minimization of (\ref{eq:omegaxi}) yields 

\be
-\frac{l_{eq}}{\xi_b}=(1+\frac{\omega}{2})(\ln\epsilon_w-3\ln(l_{eq}/\xi_b)); \hspace*{2mm}\delta\mu=0^+
\label{eq:leqrg}
\ee
as $\epsilon_w\to 0^+$. The equilibrium thickness still diverges with
the MF form (\ref{eq:leq}) but the amplitude is increased by a factor
$(1+\omega/2)$. MF is recovered when the interface becomes very stiff
so that $\omega\to 0$. The parallel correlation length can be obtained
from either $\xi_\parallel^{-2}\propto
\left(\frac{\partial^2\omega_B(l)}{\partial l^2}\right)$ at $l=l_{eq}$
or from $\xi_\parallel^2\propto \left(\frac{\partial
  l_{eq}}{\partial\mu}\right)$. In both cases we find

\be
\xi_\parallel\sim\epsilon_w^{-1/2}[(1+\frac{w}{2})(-\ln\epsilon_w)]^{3/2};\hspace*{2mm}\delta\mu=0^+
\ee

The singular part of the surface excess free energy can be calculated from (\ref{eq:omegaxi}) and we obtain 
\be
1+\cos(\theta)\sim\epsilon_w(-(1+\frac{\omega}{2})\ln\epsilon_w)^{-2}; \hspace*{2mm}\delta\mu=0^+
\ee
Once again only the amplitudes are changed from the MF results (\ref{eq:xipar}) and (\ref{eq:costheta}). Note that (\ref{eq:leqrg}) is reminiscent of the result for complete drying from off-coexistence for SR forces, e.g. at a planar hard-wall. There the second term in (\ref{eq:omegaxi}) is absent but the third remains leading to 
\be
-\frac{l_{eq}}{\xi_b}=(1+\frac{\omega}{2}) \ln\delta\mu,  {\rm as } ~\delta\mu\to 0^+
\ee                                      

Unlike the case of SR forces considered by Brezin et
al.~\cite{Brezin:1983dn} and in many subsequent studies,
e.g. \cite{Binder:1989aa,Albano:2012db,Parry:2008ef,Parry:2008rp,Parry:2009aa}
where several of the critical exponents are predicted to depend
explicitly on the parameter $\omega$, for the binding potential
(\ref{eq:bpot}) our RG analysis predicts the critical exponents to be
unchanged from their MF values and therefore independent of $\omega$
even though the upper critical dimension is also $d=3$. We note that
the conclusions of the MF and RG analyzes are changed little if we
consider LR wall-fluid potentials other than the standard $9$-$3$ case
(\ref{eq:LRpot}). Suppose the leading power-law decay is proportional
to $-(\sigma/z)^p$, with $p>2$. Then the coefficient of the logarithm
in (\ref{eq:leq}) is replaced by $(p+1)$, (\ref{eq:logchi}) is
unchanged, and the power of the logarithm in (\ref{eq:xipar}) and
(\ref{eq:costheta}) is replaced by $(p+1)/2$ and $-p$, respectively. 
The RG results are changed accordingly.

\vspace*{-2mm}
\subsection{Details of DFT calculations}
\vspace*{-2mm}

The classical DFT that we employ is that used in a previous study of
solvophobic substrates but one that did not address critical drying
\cite{Evans:2015aa}. The excess Helmholtz free energy functional is
approximated by the sum of a hard-sphere functional, treated by means
of Rosenfeld's fundamental measure theory, and a standard MF treatment
of attractive fluid-fluid interactions. Eq.(14) of Ref.~\cite{Evans:2015aa} displays
the grand potential functional. In the present study the attractive
part of the truncated LJ potential is given by

\be
\phi_{\rm att}(r)=\left \{ \begin{array}{ll}
-\epsilon_{LJ}, \mbox{\hspace{4mm}}   &  r<r_{\rm min}  \\
 4\epsilon_{LJ}\left[\left(\frac{\sigma}{r}\right)^{12}-\left(\frac{\sigma}{r}\right)^{6}\right], & r_{\rm min}<r<r_c \, ,\\
0, & r>r_c,\\
\end{array}
\right.
\label{eq:DFTpot}
\ee
where $r_{\rm min}=2^{1/6}\sigma$. The potential is truncated at $r_c
=2.5 \sigma$, as in simulation. The critical temperature is given by
$k_BT_c=1.3194\epsilon_{LJ}$ and calculations are performed at $T
=0.775T_c$. The wall-fluid potential is the standard $9$-$3$ model:

\be
W_{\rm 93}(z)= \epsilon_w\epsilon_{LJ}\left[\frac{2}{15}\left(\frac{\sigma}{z}\right)^9-\left(\frac{\sigma}{z}\right)^3\right], 
\label{eq:9-3pot}
\ee
where, once again,  $\epsilon_w$ is a dimensionless measure of the ratio of the
wall-fluid attraction compared to that of fluid-fluid. Note that (\ref{eq:9-3pot})
differs slightly from (\ref{eq:LRpot}). However, the crucial $z^{-3}$ tail
contribution has the same coefficient.  The hard-sphere diameter, entering the functional, is $d=\sigma$.

In the calculations we determine equilibrium density profiles
$\rho(z)$ and the surface tensions
$\gamma_{lv},\gamma_{wl},\gamma_{wv}$, by minimizing the grand
potential functional \cite{Evans:2015aa}. The local compressibility
$\chi(z)$ is determined numerically as described in
\cite{Evans:2015aa}. We have performed calculations for a single wall
and for a pair of confining walls, equivalent to the GCMC
simulations. In this Letter we show results for the single wall only.

Key results are shown in Fig.~S1 below. Here we plot $\rho(z)$ and
$\chi(z)$ for very small values of $\epsilon_w$.  As $\epsilon_w$ is
reduced towards zero the thickness of the drying film $l_{eq}$
increases (panel 1). We have confirmed in detail, within DFT, that the
Gibbs adsorption or $l_{eq}$ grows according to (\ref{eq:leq}).  The
position of the peak in $\chi(z)$ shifts with the position of the
gas-liquid interface and its height increases very rapidly as
$\epsilon_w \to 0^+$ (panel 2). Panel 3 shows clearly that
$\ln\chi(l_{eq})$ increases linearly with $l_{eq}$. The prediction
(\ref{eq:logchi}), including the correct prefactor, the inverse bulk
correlation length, is confirmed by our DFT calculations. We determine
the contact angle via Young's equation and DFT results for $\cos(\theta)$
are shown in the right panel of Fig.~\ref{fig:costhetacompare} (main text). For the LR case (\ref{eq:9-3pot})
we find critical drying at $\epsilon_w =0$ and 1st order wetting at a value
of $\epsilon_w$ that is smaller than in simulation. For the SR case
(square-well), where again $d=\sigma$, both drying and wetting are
critical transitions, as in simulation. However, the separation in
$\epsilon_w$ between wetting and drying in DFT is smaller than in
simulation. In summary the microscopic DFT results for a single wall
are in complete agreement with those from the simple binding potential
treatment, based on (\ref{eq:bpot}), and described above.

\begin{figure}[h]
\centerline{\includegraphics[width=8cm,clip=true]{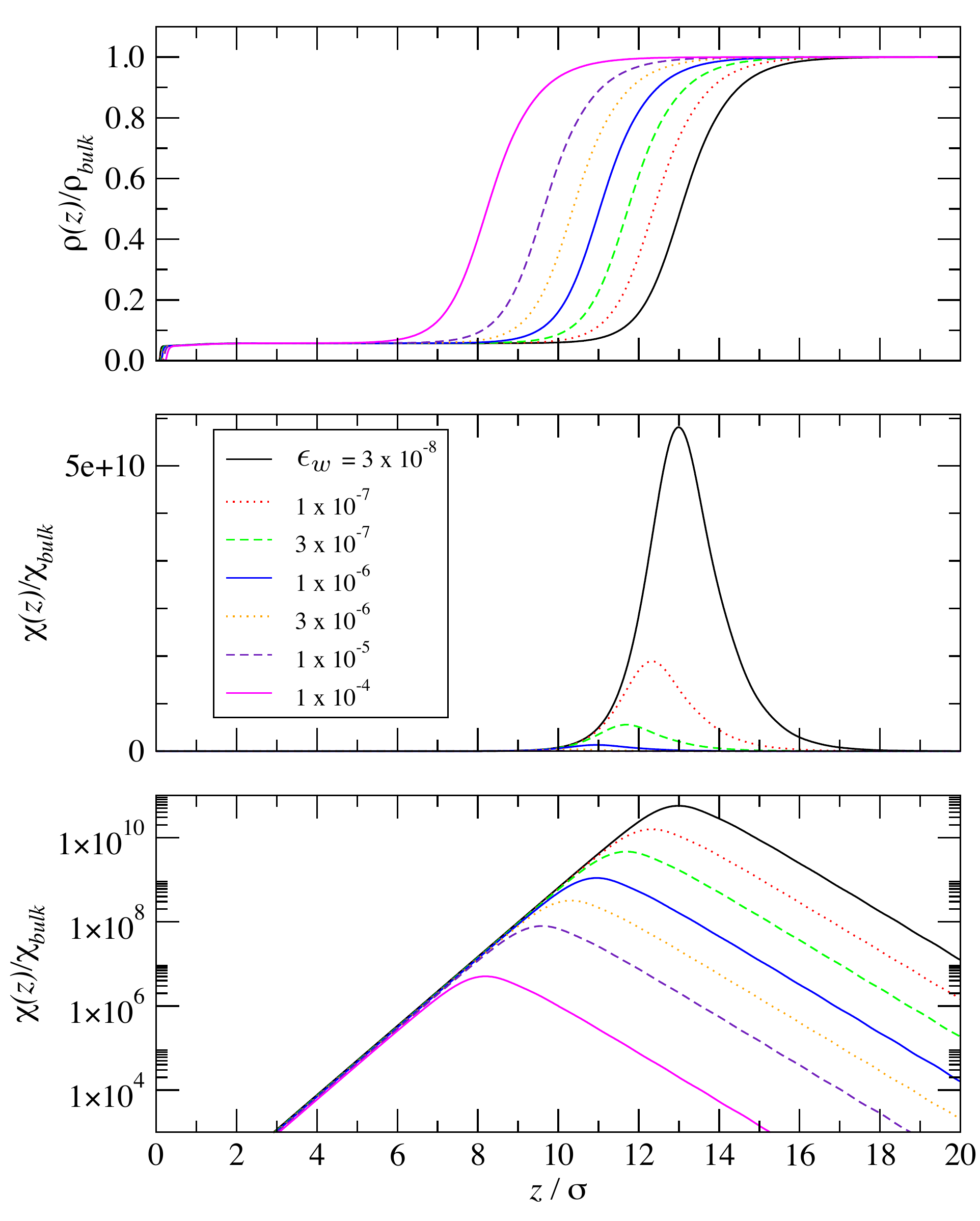}}
\caption{DFT results for the normalised density profiles $\rho(z)/\rho_b$
  (top panel) and the local compressibilities $\chi(z)/\chi_b$ (linear scale
  - middle panel; log-scale - bottom panel) for the fluid at single
  walls. The strength of the wall-fluid interaction potentials are
  given in the caption. The temperature is $T = 0.775T_c$ and the
  reservoir is at bulk liquid-gas coexistence, on the liquid side
  $\delta\mu = 0^+$.}
\label{fig:mariafig}
\end{figure}

\vspace*{-2mm}
\subsection{$L$-dependence of the high density tail of $p(\rho)$}
\vspace*{-2mm}

The high density tail in $p(\rho)$ corresponds to the free energy cost
of pushing the liquid-slab up against the hard wall.  A clear feature
of Fig.~\ref{fig:LRFS} is that the density at which the tail occurs shifts strongly to lower
values as the wall area $L^2$ increases.  One complication in
comparing the distributions for various $L$ to explain this shift is that sampling of
$p(\rho)$ is truncated at low densities due to the need to avoid the
region of capillary evaporation.  Consequently it is not possible to
normalize the distribution. To deal with this (and to make the tails
more visible) let us instead consider the logarithm of $p(\rho)$ which
is the negative of the (grand) free energy function $F(\rho)=-k_BT\ln
p(\rho)$. Doing so changes the unknown normalization factor into an
additive constant, which can be removed by differentiation.  The
derivative $\partial \ln p(\rho)/\partial\rho$, is closely related to
$\mu(N)=-k_BT (\partial \ln p(N)/\partial N)$, the chemical potential
function. A plot of this derivative is given in
Fig.~S2. We find that the curves all scale onto one
another with a simple $L^{-2}$ scaling of the ordinate, no scaling is
needed for the density. In view of this, we can write

\be
\frac{\partial \ln p(\rho)}{\partial\rho}=L^2g(\rho)
\ee
where $g(\rho)$ is some function of $\rho$. It follows that the free energy itself scales like

\be
-\ln p(\rho)=-L^2\int_0^\rho g(\rho^\prime)d\rho^\prime
\ee
i.e. it has a rather trivial $L^2$ scaling. 

\begin{figure}[h]
\centerline{\includegraphics[width=7.5cm,clip=true]{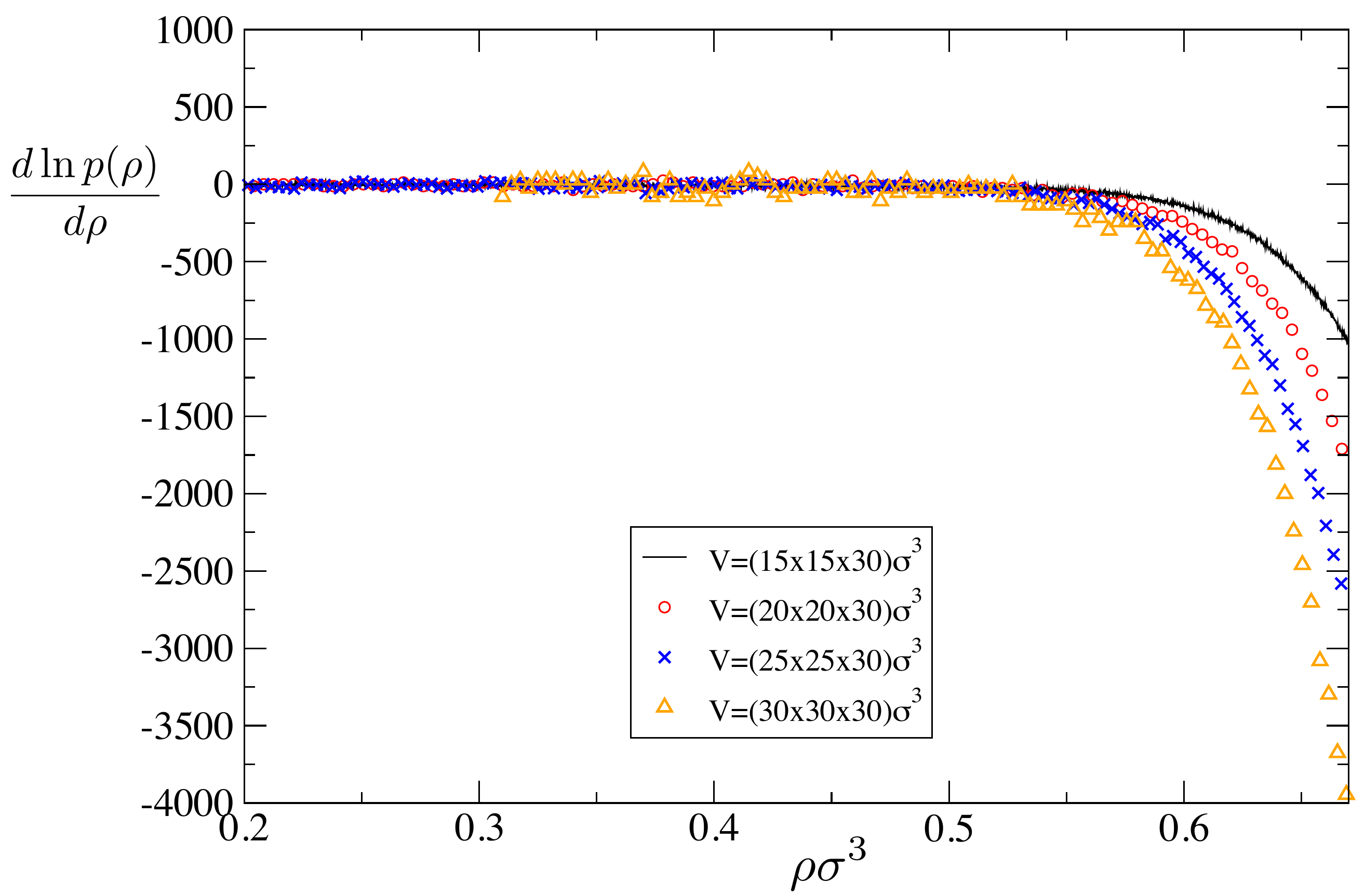}}
\centerline{\includegraphics[width=7.5cm,clip=true]{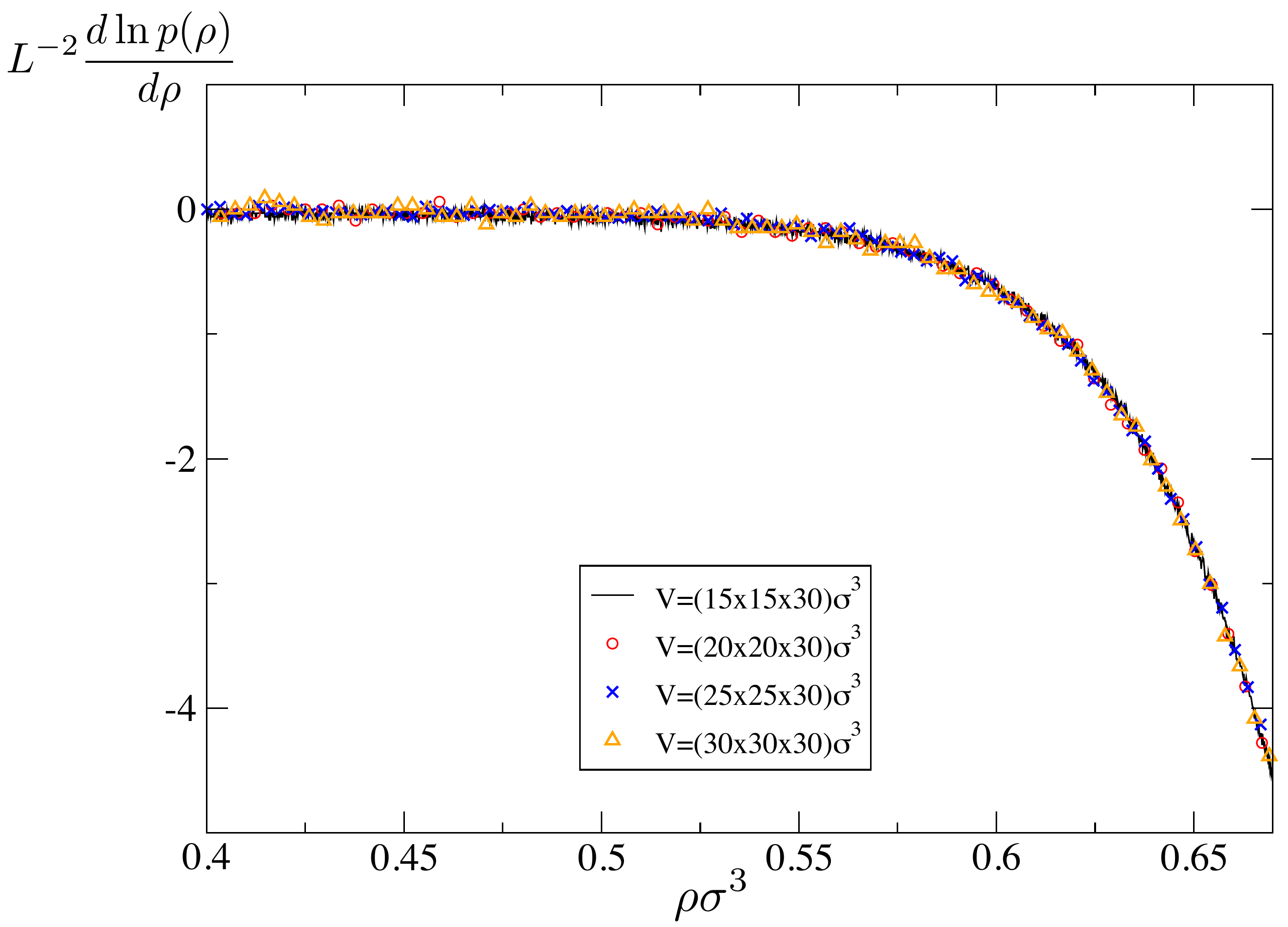}}

\caption{{\bf (a)} $ \partial \ln p(\rho)/\partial \rho$ for
  $L=15\sigma,20\sigma,25\sigma,30\sigma$ at $\epsilon=0$ and
  $T=0.775T_c$. {\bf (b)} The same data scaled by $L^{-2}$}

\label{fig:dender}
\end{figure}

An appealing rationalization of this finding is in mechanical terms.
The configuration takes the form of a liquid slab. For the density to
grow, the slab has to thicken, i.e the slab interface has to approach
the hard walls.  Thus the density is linearly related to the average
separation of the slab surface from the walls ie. $\rho \propto z$. From
this viewpoint, $\partial \ln p(\rho)/\partial\rho$ is a force profile
and $g(\rho)$ is a pressure profile. The $L^2$ scaling then suggest
that the hard wall exerts a repulsive pressure on the liquid-vapor
interface which is independent of $L$, so the force (and hence the
work) required to push the interface to the walls increases like
$L^2$. The $L^2$ scaling of the free energy leads to the apparent
shift in the tail position in $p(\rho)$.

\subsection{Movie of the emergent liquid-vapor interface near critical drying}
\vspace*{-1mm}

This movie (from which the snapshot of fig.~4 was taken) allows a
clearer view of the configurational structure that occurs near
critical drying. The movie focuses on the region near the wall at
$z=0$ for a system of size $L=40\sigma$. The temperature is
$T=0.775T_c$ and the attractive wall strength is $\epsilon=0.2$ which
is slightly larger than that for which the liquid peak in $p(\rho)$
vanishes for this $L$.  Observing the purple shaded particles lying
close to the wall we note that there is a large but finite
$\xi_\parallel$ manifest in the large fractal bubbles of `vapor' which
almost span the system in the lateral dimension. However, the
perpendicular extent of these bubbles is microscopic, extending only a
few particle diameters away from the wall.
\url{<http://people.bath.ac.uk/pysnbw/sm_movie.mp4>}


\begin{thebibliography}{47}%
\makeatletter
\providecommand \@ifxundefined [1]{%
 \@ifx{#1\undefined}
}%
\providecommand \@ifnum [1]{%
 \ifnum #1\expandafter \@firstoftwo
 \else \expandafter \@secondoftwo
 \fi
}%
\providecommand \@ifx [1]{%
 \ifx #1\expandafter \@firstoftwo
 \else \expandafter \@secondoftwo
 \fi
}%
\providecommand \natexlab [1]{#1}%
\providecommand \enquote  [1]{``#1''}%
\providecommand \bibnamefont  [1]{#1}%
\providecommand \bibfnamefont [1]{#1}%
\providecommand \citenamefont [1]{#1}%
\providecommand \href@noop [0]{\@secondoftwo}%
\providecommand \href [0]{\begingroup \@sanitize@url \@href}%
\providecommand \@href[1]{\@@startlink{#1}\@@href}%
\providecommand \@@href[1]{\endgroup#1\@@endlink}%
\providecommand \@sanitize@url [0]{\catcode `\\12\catcode `\$12\catcode
  `\&12\catcode `\#12\catcode `\^12\catcode `\_12\catcode `\%12\relax}%
\providecommand \@@startlink[1]{}%
\providecommand \@@endlink[0]{}%
\providecommand \url  [0]{\begingroup\@sanitize@url \@url }%
\providecommand \@url [1]{\endgroup\@href {#1}{\urlprefix }}%
\providecommand \urlprefix  [0]{URL }%
\providecommand \Eprint [0]{\href }%
\providecommand \doibase [0]{http://dx.doi.org/}%
\providecommand \selectlanguage [0]{\@gobble}%
\providecommand \bibinfo  [0]{\@secondoftwo}%
\providecommand \bibfield  [0]{\@secondoftwo}%
\providecommand \translation [1]{[#1]}%
\providecommand \BibitemOpen [0]{}%
\providecommand \bibitemStop [0]{}%
\providecommand \bibitemNoStop [0]{.\EOS\space}%
\providecommand \EOS [0]{\spacefactor3000\relax}%
\providecommand \BibitemShut  [1]{\csname bibitem#1\endcsname}%
\let\auto@bib@innerbib\@empty
\bibitem [{\citenamefont {Simpson}\ \emph {et~al.}(2015)\citenamefont
  {Simpson}, \citenamefont {Hunter},\ and\ \citenamefont
  {Aytug}}]{Simpson:2015le}%
  \BibitemOpen
  \bibfield  {author} {\bibinfo {author} {\bibfnamefont {J.~T.}\ \bibnamefont
  {Simpson}}, \bibinfo {author} {\bibfnamefont {S.~R.}\ \bibnamefont {Hunter}},
  \ and\ \bibinfo {author} {\bibfnamefont {T.}~\bibnamefont {Aytug}},\ }\href
  {http://stacks.iop.org/0034-4885/78/i=8/a=086501} {\bibfield  {journal}
  {\bibinfo  {journal} {Rep. Prog. Phys.}\ }\textbf {\bibinfo {volume} {78}},\
  \bibinfo {pages} {086501} (\bibinfo {year} {2015})}\BibitemShut {NoStop}%
\bibitem [{\citenamefont {Li}\ \emph {et~al.}(2007)\citenamefont {Li},
  \citenamefont {Reinhoudt},\ and\ \citenamefont {Crego-Calama}}]{Li:2007aa}%
  \BibitemOpen
  \bibfield  {author} {\bibinfo {author} {\bibfnamefont {X.-M.}\ \bibnamefont
  {Li}}, \bibinfo {author} {\bibfnamefont {D.}~\bibnamefont {Reinhoudt}}, \
  and\ \bibinfo {author} {\bibfnamefont {M.}~\bibnamefont {Crego-Calama}},\
  }\href {\doibase 10.1039/B602486F} {\bibfield  {journal} {\bibinfo  {journal}
  {Chem. Soc. Rev.}\ }\textbf {\bibinfo {volume} {36}},\ \bibinfo {pages}
  {1350} (\bibinfo {year} {2007})}\BibitemShut {NoStop}%
\bibitem [{\citenamefont {Ueda}\ and\ \citenamefont
  {Levkin}(2013)}]{Ueda:2013aa}%
  \BibitemOpen
  \bibfield  {author} {\bibinfo {author} {\bibfnamefont {E.}~\bibnamefont
  {Ueda}}\ and\ \bibinfo {author} {\bibfnamefont {P.~A.}\ \bibnamefont
  {Levkin}},\ }\href {\doibase 10.1002/adma.201204120} {\bibfield  {journal}
  {\bibinfo  {journal} {Adv. Mater.}\ }\textbf {\bibinfo {volume} {25}},\
  \bibinfo {pages} {1234} (\bibinfo {year} {2013})}\BibitemShut {NoStop}%
\bibitem{Checco:2014aa} A. Checco, B.~M. Ocko, A. Rahman, C.~T. Black, M. Tasinkevych, A. Giacomello, and S. Dietrich, Phys. Rev. Lett. {\bf 112}, 216101 (2014). 
\bibitem [{\citenamefont {Bonn}\ \emph {et~al.}(2009)\citenamefont {Bonn},
  \citenamefont {Eggers}, \citenamefont {Indekeu}, \citenamefont {Meunier},\
  and\ \citenamefont {Rolley}}]{Bonn:2009if}%
  \BibitemOpen
  \bibfield  {author} {\bibinfo {author} {\bibfnamefont {D.}~\bibnamefont
  {Bonn}}, \bibinfo {author} {\bibfnamefont {J.}~\bibnamefont {Eggers}},
  \bibinfo {author} {\bibfnamefont {J.}~\bibnamefont {Indekeu}}, \bibinfo
  {author} {\bibfnamefont {J.}~\bibnamefont {Meunier}}, \ and\ \bibinfo
  {author} {\bibfnamefont {E.}~\bibnamefont {Rolley}},\ }\href {\doibase
  10.1103/RevModPhys.81.739} {\bibfield  {journal} {\bibinfo  {journal} {Rev.
  Mod. Phys.}\ }\textbf {\bibinfo {volume} {81}},\ \bibinfo {pages} {739}
  (\bibinfo {year} {2009})}\BibitemShut {NoStop}%
\bibitem [{\citenamefont {Friedman}\ \emph {et~al.}(2013)\citenamefont
  {Friedman}, \citenamefont {Khalil},\ and\ \citenamefont
  {Taborek}}]{Friedman:2013nx}%
  \BibitemOpen
  \bibfield  {author} {\bibinfo {author} {\bibfnamefont {S.~R.}\ \bibnamefont
  {Friedman}}, \bibinfo {author} {\bibfnamefont {M.}~\bibnamefont {Khalil}}, \
  and\ \bibinfo {author} {\bibfnamefont {P.}~\bibnamefont {Taborek}},\ }\href
  {\doibase 10.1103/PhysRevLett.111.226101} {\bibfield  {journal} {\bibinfo
  {journal} {Phys. Rev. Lett.}\ }\textbf {\bibinfo {volume} {111}},\ \bibinfo
  {pages} {226101} (\bibinfo {year} {2013})}\BibitemShut {NoStop}%
\bibitem [{\citenamefont {Binder}\ \emph {et~al.}(1989)\citenamefont {Binder},
  \citenamefont {Landau},\ and\ \citenamefont {Wansleben}}]{Binder:1989aa}%
  \BibitemOpen
  \bibfield  {author} {\bibinfo {author} {\bibfnamefont {K.}~\bibnamefont
  {Binder}}, \bibinfo {author} {\bibfnamefont {D.~P.}\ \bibnamefont {Landau}},
  \ and\ \bibinfo {author} {\bibfnamefont {S.}~\bibnamefont {Wansleben}},\
  }\href {\doibase 10.1103/PhysRevB.40.6971} {\bibfield  {journal} {\bibinfo
  {journal} {Phys. Rev. B}\ }\textbf {\bibinfo {volume} {40}},\ \bibinfo
  {pages} {6971} (\bibinfo {year} {1989})}\BibitemShut {NoStop}%
\bibitem [{\citenamefont {Albano}\ and\ \citenamefont
  {Binder}(2012)}]{Albano:2012db}%
  \BibitemOpen
  \bibfield  {author} {\bibinfo {author} {\bibfnamefont {E.~V.}\ \bibnamefont
  {Albano}}\ and\ \bibinfo {author} {\bibfnamefont {K.}~\bibnamefont
  {Binder}},\ }\href {\doibase 10.1103/PhysRevLett.109.036101} {\bibfield
  {journal} {\bibinfo  {journal} {Phys. Rev. Lett.}\ }\textbf {\bibinfo
  {volume} {109}},\ \bibinfo {pages} {036101} (\bibinfo {year}
  {2012})}\BibitemShut {NoStop}%
\bibitem [{\citenamefont {Bryk}\ and\ \citenamefont
  {Binder}(2013)}]{Bryk:2013pi}%
  \BibitemOpen
  \bibfield  {author} {\bibinfo {author} {\bibfnamefont {P.}~\bibnamefont
  {Bryk}}\ and\ \bibinfo {author} {\bibfnamefont {K.}~\bibnamefont {Binder}},\
  }\href {\doibase 10.1103/PhysRevE.88.030401} {\bibfield  {journal} {\bibinfo
  {journal} {Phys. Rev. E}\ }\textbf {\bibinfo {volume} {88}},\ \bibinfo
  {pages} {030401} (\bibinfo {year} {2013})}\BibitemShut {NoStop}%
\bibitem [{\citenamefont {van Swol}\ and\ \citenamefont
  {Henderson}(1989)}]{Swol:1989by}%
  \BibitemOpen
  \bibfield  {author} {\bibinfo {author} {\bibfnamefont {F.}~\bibnamefont {van
  Swol}}\ and\ \bibinfo {author} {\bibfnamefont {J.~R.}\ \bibnamefont
  {Henderson}},\ }\href {\doibase 10.1103/PhysRevA.40.2567} {\bibfield
  {journal} {\bibinfo  {journal} {Phys. Rev. A}\ }\textbf {\bibinfo {volume}
  {40}},\ \bibinfo {pages} {2567} (\bibinfo {year} {1989})}\BibitemShut
  {NoStop}%
\bibitem [{\citenamefont {Henderson}\ and\ \citenamefont {van
  Swol}(1990)}]{Henderson:1990nq}%
  \BibitemOpen
  \bibfield  {author} {\bibinfo {author} {\bibfnamefont {J.~R.}\ \bibnamefont
  {Henderson}}\ and\ \bibinfo {author} {\bibfnamefont {F.}~\bibnamefont {van
  Swol}},\ }\href {http://stacks.iop.org/0953-8984/2/i=19/a=020} {\bibfield
  {journal} {\bibinfo  {journal} {J. Phys: Condens. Matter}\ }\textbf {\bibinfo
  {volume} {2}},\ \bibinfo {pages} {4537} (\bibinfo {year} {1990})}\BibitemShut
  {NoStop}%
\bibitem [{\citenamefont {van Swol}\ and\ \citenamefont
  {Henderson}(1991)}]{Swol:1991fq}%
  \BibitemOpen
  \bibfield  {author} {\bibinfo {author} {\bibfnamefont {F.}~\bibnamefont {van
  Swol}}\ and\ \bibinfo {author} {\bibfnamefont {J.~R.}\ \bibnamefont
  {Henderson}},\ }\href {\doibase 10.1103/PhysRevA.43.2932} {\bibfield
  {journal} {\bibinfo  {journal} {Phys. Rev. A}\ }\textbf {\bibinfo {volume}
  {43}},\ \bibinfo {pages} {2932} (\bibinfo {year} {1991})}\BibitemShut
  {NoStop}%
\bibitem [{\citenamefont {Nijmeijer}\ \emph {et~al.}(1992)\citenamefont
  {Nijmeijer}, \citenamefont {Bruin}, \citenamefont {Bakker},\ and\
  \citenamefont {van Leeuwen}}]{Nijmeijer:1992fk}%
  \BibitemOpen
  \bibfield  {author} {\bibinfo {author} {\bibfnamefont {M.~J.~P.}\
  \bibnamefont {Nijmeijer}}, \bibinfo {author} {\bibfnamefont {C.}~\bibnamefont
  {Bruin}}, \bibinfo {author} {\bibfnamefont {A.~F.}\ \bibnamefont {Bakker}}, \
  and\ \bibinfo {author} {\bibfnamefont {J.~M.~J.}\ \bibnamefont {van
  Leeuwen}},\ }\href {http://stacks.iop.org/0953-8984/4/i=1/a=012} {\bibfield
  {journal} {\bibinfo  {journal} {J. Phys: Condens. Matter}\ }\textbf {\bibinfo
  {volume} {4}},\ \bibinfo {pages} {15} (\bibinfo {year} {1992})}\BibitemShut
  {NoStop}%
\bibitem [{\citenamefont {Nijmeijer}\ \emph {et~al.}(1991)\citenamefont
  {Nijmeijer}, \citenamefont {Bruin}, \citenamefont {Bakker},\ and\
  \citenamefont {van Leeuwen}}]{Nijmeijer:1991sw}%
  \BibitemOpen
  \bibfield  {author} {\bibinfo {author} {\bibfnamefont {M.~J.~P.}\
  \bibnamefont {Nijmeijer}}, \bibinfo {author} {\bibfnamefont {C.}~\bibnamefont
  {Bruin}}, \bibinfo {author} {\bibfnamefont {A.~F.}\ \bibnamefont {Bakker}}, \
  and\ \bibinfo {author} {\bibfnamefont {J.~M.~J.}\ \bibnamefont {van
  Leeuwen}},\ }\href {\doibase 10.1103/PhysRevB.44.834} {\bibfield  {journal}
  {\bibinfo  {journal} {Phys. Rev. B}\ }\textbf {\bibinfo {volume} {44}},\
  \bibinfo {pages} {834} (\bibinfo {year} {1991})}\BibitemShut {NoStop}%
\bibitem [{\citenamefont {Henderson}\ \emph {et~al.}(1992)\citenamefont
  {Henderson}, \citenamefont {Tarazona}, \citenamefont {van Swol},\ and\
  \citenamefont {Velasco}}]{Henderson:1992kk}%
  \BibitemOpen
  \bibfield  {author} {\bibinfo {author} {\bibfnamefont {J.~R.}\ \bibnamefont
  {Henderson}}, \bibinfo {author} {\bibfnamefont {P.}~\bibnamefont {Tarazona}},
  \bibinfo {author} {\bibfnamefont {F.}~\bibnamefont {van Swol}}, \ and\
  \bibinfo {author} {\bibfnamefont {E.}~\bibnamefont {Velasco}},\ }\href
  {http://scitation.aip.org/content/aip/journal/jcp/96/6/10.1063/1.462799}
  {\bibfield  {journal} {\bibinfo  {journal} {J. Chem. Phys.}\ }\textbf
  {\bibinfo {volume} {96}},\ \bibinfo {pages} {4633} (\bibinfo {year}
  {1992})}\BibitemShut {NoStop}%
\bibitem [{\citenamefont {Bruin}\ \emph {et~al.}(1995)\citenamefont {Bruin},
  \citenamefont {Nijmeijer},\ and\ \citenamefont {Crevecoeur}}]{Bruin:1995ud}%
  \BibitemOpen
  \bibfield  {author} {\bibinfo {author} {\bibfnamefont {C.}~\bibnamefont
  {Bruin}}, \bibinfo {author} {\bibfnamefont {M.~J.~P.}\ \bibnamefont
  {Nijmeijer}}, \ and\ \bibinfo {author} {\bibfnamefont {R.~M.}\ \bibnamefont
  {Crevecoeur}},\ }\href@noop {} {\bibfield  {journal} {\bibinfo  {journal} {J.
  Chem. Phys.}\ }\textbf {\bibinfo {volume} {102}},\ \bibinfo {pages} {7622}
  (\bibinfo {year} {1995})}\BibitemShut {NoStop}%
\bibitem [{\citenamefont {Ancilotto}\ \emph {et~al.}(2001)\citenamefont
  {Ancilotto}, \citenamefont {Curtarolo}, \citenamefont {Toigo},\ and\
  \citenamefont {Cole}}]{Ancilotto:2001uq}%
  \BibitemOpen
  \bibfield  {author} {\bibinfo {author} {\bibfnamefont {F.}~\bibnamefont
  {Ancilotto}}, \bibinfo {author} {\bibfnamefont {S.}~\bibnamefont
  {Curtarolo}}, \bibinfo {author} {\bibfnamefont {F.}~\bibnamefont {Toigo}}, \
  and\ \bibinfo {author} {\bibfnamefont {M.~W.}\ \bibnamefont {Cole}},\ }\href
  {\doibase 10.1103/PhysRevLett.87.206103} {\bibfield  {journal} {\bibinfo
  {journal} {Phys. Rev. Lett.}\ }\textbf {\bibinfo {volume} {87}},\ \bibinfo
  {pages} {206103} (\bibinfo {year} {2001})}\BibitemShut {NoStop}%
\bibitem [{\citenamefont {Oleinikova}\ \emph {et~al.}(2005)\citenamefont
  {Oleinikova}, \citenamefont {Brovchenko},\ and\ \citenamefont
  {Geiger}}]{Oleinikova:2005uo}%
  \BibitemOpen
  \bibfield  {author} {\bibinfo {author} {\bibfnamefont {A.}~\bibnamefont
  {Oleinikova}}, \bibinfo {author} {\bibfnamefont {I.}~\bibnamefont
  {Brovchenko}}, \ and\ \bibinfo {author} {\bibfnamefont {A.}~\bibnamefont
  {Geiger}},\ }\href {http://stacks.iop.org/0953-8984/17/i=50/a=006} {\bibfield
   {journal} {\bibinfo  {journal} {J. Phys.: Condens. Matter}\ }\textbf
  {\bibinfo {volume} {17}},\ \bibinfo {pages} {7845} (\bibinfo {year}
  {2005})}\BibitemShut {NoStop}%
\bibitem [{\citenamefont {Evans}\ \emph {et~al.}()\citenamefont {Evans},
  \citenamefont {Stewart},\ and\ \citenamefont {Wilding}}]{Evans:SM2016}%
  \BibitemOpen
  \bibfield  {author} {\bibinfo {author} {\bibfnamefont {R.}~\bibnamefont
  {Evans}}, \bibinfo {author} {\bibfnamefont {M.}~\bibnamefont {Stewart}}, \
  and\ \bibinfo {author} {\bibfnamefont {N.}~\bibnamefont {Wilding}},\
  }\href@noop {} {}\bibinfo {note} {Supplementary material which also includes
  \protect\cite{Nightingale:1983ty,Brezin:1983dn} provides a) Details of
  Binding Potential calculations; b) Details of DFT calculation; c) Details of
  how the tails of $p(\rho)$ scale with the wall dimension $L$; d) A movie of
  the near critical interface.}\BibitemShut {Stop}%
\bibitem [{\citenamefont {Evans}(1992)}]{Evans:1992jo}%
  \BibitemOpen
  \bibfield  {author} {\bibinfo {author} {\bibfnamefont {R.}~\bibnamefont
  {Evans}},\ }in\ \href@noop {} {\emph {\bibinfo {booktitle} {Fundamentals of
  Inhomogeneous Fluids}}},\ \bibinfo {editor} {edited by\ \bibinfo {editor}
  {\bibfnamefont {D.}~\bibnamefont {Henderson}}}\ (\bibinfo  {publisher}
  {Dekker},\ \bibinfo {year} {1992})\ p.~\bibinfo {pages} {85}\BibitemShut
  {NoStop}%
\bibitem [{\citenamefont {Evans}\ and\ \citenamefont
  {Stewart}(2015)}]{Evans:2015aa}%
  \BibitemOpen
  \bibfield  {author} {\bibinfo {author} {\bibfnamefont {R.}~\bibnamefont
  {Evans}}\ and\ \bibinfo {author} {\bibfnamefont {M.~C.}\ \bibnamefont
  {Stewart}},\ }\href@noop {} {\bibfield  {journal} {\bibinfo  {journal}
  {J.~Phys.: Condens. Matt.}\ }\textbf {\bibinfo {volume} {27}},\ \bibinfo
  {pages} {194111} (\bibinfo {year} {2015})}\BibitemShut {NoStop}%
\bibitem [{\citenamefont {Berg}\ and\ \citenamefont
  {Neuhaus}(1992)}]{berg1992}%
  \BibitemOpen
  \bibfield  {author} {\bibinfo {author} {\bibfnamefont {B.~A.}\ \bibnamefont
  {Berg}}\ and\ \bibinfo {author} {\bibfnamefont {T.}~\bibnamefont {Neuhaus}},\
  }\href@noop {} {\bibfield  {journal} {\bibinfo  {journal} {Phys. Rev. Lett.}\
  }\textbf {\bibinfo {volume} {68}},\ \bibinfo {pages} {9} (\bibinfo {year}
  {1992})}\BibitemShut {NoStop}%
\bibitem [{\citenamefont {Wilding}(1995)}]{Wilding1995}%
  \BibitemOpen
  \bibfield  {author} {\bibinfo {author} {\bibfnamefont {N.~B.}\ \bibnamefont
  {Wilding}},\ }\href {\doibase 10.1103/PhysRevE.52.602} {\bibfield  {journal}
  {\bibinfo  {journal} {Phys. Rev. E}\ }\textbf {\bibinfo {volume} {52}},\
  \bibinfo {pages} {602} (\bibinfo {year} {1995})}\BibitemShut {NoStop}%
\bibitem [{\citenamefont {Wilding}()}]{Wilding:2016qr}%
  \BibitemOpen
  \bibfield  {author} {\bibinfo {author} {\bibfnamefont {N.~B.}\ \bibnamefont
  {Wilding}},\ }\href {\doibase doi:10.1088/0953-8984/28/41/414016} {\bibfield  {journal}
  {\bibinfo  {journal} {J. Phys. Condens. Matter}\ }\textbf {\bibinfo {volume} {28}},\
  \bibinfo {pages} {414016} (\bibinfo {year} {2016})}\BibitemShut {NoStop}%
\bibitem [{\citenamefont {MacDowell}\ \emph {et~al.}(2004)\citenamefont
  {MacDowell}, \citenamefont {Virnau}, \citenamefont {M{\"u}ller},\ and\
  \citenamefont {Binder}}]{MacDowell:2004wj}%
  \BibitemOpen
  \bibfield  {author} {\bibinfo {author} {\bibfnamefont {L.~G.}\ \bibnamefont
  {MacDowell}}, \bibinfo {author} {\bibfnamefont {P.}~\bibnamefont {Virnau}},
  \bibinfo {author} {\bibfnamefont {M.}~\bibnamefont {M{\"u}ller}}, \ and\
  \bibinfo {author} {\bibfnamefont {K.}~\bibnamefont {Binder}},\ }\href@noop {}
  {\bibfield  {journal} {\bibinfo  {journal} {J. Chem. Phys.}\ }\textbf
  {\bibinfo {volume} {120}},\ \bibinfo {pages} {5293} (\bibinfo {year}
  {2004})}\BibitemShut {NoStop}%
\bibitem [{\citenamefont {M{\"u}ller}\ and\ \citenamefont
  {MacDowell}(2000)}]{Muller:2000fv}%
  \BibitemOpen
  \bibfield  {author} {\bibinfo {author} {\bibfnamefont {M.}~\bibnamefont
  {M{\"u}ller}}\ and\ \bibinfo {author} {\bibfnamefont {L.~G.}\ \bibnamefont
  {MacDowell}},\ }\href@noop {} {\bibfield  {journal} {\bibinfo  {journal}
  {Macromolecules}\ }\textbf {\bibinfo {volume} {33}},\ \bibinfo {pages} {3902}
  (\bibinfo {year} {2000})}\BibitemShut {NoStop}%
\bibitem [{\citenamefont {Evans}\ and\ \citenamefont
  {Wilding}(2015)}]{Evans:2015wo}%
  \BibitemOpen
  \bibfield  {author} {\bibinfo {author} {\bibfnamefont {R.}~\bibnamefont
  {Evans}}\ and\ \bibinfo {author} {\bibfnamefont {N.~B.}\ \bibnamefont
  {Wilding}},\ }\href {\doibase 10.1103/PhysRevLett.115.016103} {\bibfield
  {journal} {\bibinfo  {journal} {Phys. Rev. Lett.}\ }\textbf {\bibinfo
  {volume} {115}},\ \bibinfo {pages} {016103} (\bibinfo {year}
  {2015})}\BibitemShut {NoStop}%
\bibitem [{\citenamefont {Dietrich}(1988)}]{Dietrich:1988et}%
  \BibitemOpen
  \bibfield  {author} {\bibinfo {author} {\bibfnamefont {S.}~\bibnamefont
  {Dietrich}},\ }\href@noop {} {\emph {\bibinfo {title} {Phase Transitions and
  Critical Phenomena vol 12}}},\ edited by\ \bibinfo {editor} {\bibfnamefont
  {C.}~\bibnamefont {Domb}}\ and\ \bibinfo {editor} {\bibfnamefont {J.~L.}\
  \bibnamefont {Lebowitz}}\ (\bibinfo  {publisher} {Academic, London},\
  \bibinfo {year} {1988})\BibitemShut {NoStop}%
\bibitem [{\citenamefont {Fan}\ and\ \citenamefont
  {Monson}(1993)}]{Fan:1993aa}%
  \BibitemOpen
  \bibfield  {author} {\bibinfo {author} {\bibfnamefont {Y.}~\bibnamefont
  {Fan}}\ and\ \bibinfo {author} {\bibfnamefont {P.~A.}\ \bibnamefont
  {Monson}},\ }\href
  {http://scitation.aip.org/content/aip/journal/jcp/99/9/10.1063/1.465833}
  {\bibfield  {journal} {\bibinfo  {journal} {J. Chem. Phys.}\ }\textbf
  {\bibinfo {volume} {99}},\ \bibinfo {pages} {6897} (\bibinfo {year}
  {1993})}\BibitemShut {NoStop}%
\bibitem [{\citenamefont {Bryk}\ \emph {et~al.}(1999)\citenamefont {Bryk},
  \citenamefont {Soko{\l}owski},\ and\ \citenamefont
  {Henderson}}]{Bryk:1999aa}%
  \BibitemOpen
  \bibfield  {author} {\bibinfo {author} {\bibfnamefont {P.}~\bibnamefont
  {Bryk}}, \bibinfo {author} {\bibfnamefont {S.}~\bibnamefont {Soko{\l}owski}},
  \ and\ \bibinfo {author} {\bibfnamefont {D.}~\bibnamefont {Henderson}},\
  }\href
  {http://scitation.aip.org/content/aip/journal/jcp/110/1/10.1063/1.478078}
  {\bibfield  {journal} {\bibinfo  {journal} {J. Chem. Phys.}\ }\textbf
  {\bibinfo {volume} {110}},\ \bibinfo {pages} {15} (\bibinfo {year}
  {1999})}\BibitemShut {NoStop}%
\bibitem [{\citenamefont {Rane}\ \emph {et~al.}(2011)\citenamefont {Rane},
  \citenamefont {Kumar},\ and\ \citenamefont {Errington}}]{Rane:2011ly}%
  \BibitemOpen
  \bibfield  {author} {\bibinfo {author} {\bibfnamefont {K.~S.}\ \bibnamefont
  {Rane}}, \bibinfo {author} {\bibfnamefont {V.}~\bibnamefont {Kumar}}, \ and\
  \bibinfo {author} {\bibfnamefont {J.~R.}\ \bibnamefont {Errington}},\ }\href
  {\doibase http://dx.doi.org/10.1063/1.3668137} {\bibfield  {journal}
  {\bibinfo  {journal} {J. Chem. Phys.}\ }\textbf {\bibinfo {volume} {135}},\
  \bibinfo {eid} {234102} (\bibinfo {year} {2011})}\BibitemShut {NoStop}%
\bibitem [{\citenamefont {Kumar}\ and\ \citenamefont
  {Errington}(2013{\natexlab{a}})}]{Kumar:2013aa}%
  \BibitemOpen
  \bibfield  {author} {\bibinfo {author} {\bibfnamefont {V.}~\bibnamefont
  {Kumar}}\ and\ \bibinfo {author} {\bibfnamefont {J.~R.}\ \bibnamefont
  {Errington}},\ }\href@noop {} {\bibfield  {journal} {\bibinfo  {journal}
  {Mol. Sim.}\ }\textbf {\bibinfo {volume} {39}},\ \bibinfo {pages} {1143}
  (\bibinfo {year} {2013}{\natexlab{a}})}\BibitemShut {NoStop}%
\bibitem [{\citenamefont {Kumar}\ and\ \citenamefont
  {Errington}(2013{\natexlab{b}})}]{Kumar:2013kx}%
  \BibitemOpen
  \bibfield  {author} {\bibinfo {author} {\bibfnamefont {V.}~\bibnamefont
  {Kumar}}\ and\ \bibinfo {author} {\bibfnamefont {J.~R.}\ \bibnamefont
  {Errington}},\ }\href@noop {} {\bibfield  {journal} {\bibinfo  {journal} {J.
  Phys. Chem. C}\ }\textbf {\bibinfo {volume} {117}},\ \bibinfo {pages} {23017}
  (\bibinfo {year} {2013}{\natexlab{b}})}\BibitemShut {NoStop}%
\bibitem [{\citenamefont {Willard}\ and\ \citenamefont
  {Chandler}(2014)}]{Willard:2014aa}%
  \BibitemOpen
  \bibfield  {author} {\bibinfo {author} {\bibfnamefont {A.~P.}\ \bibnamefont
  {Willard}}\ and\ \bibinfo {author} {\bibfnamefont {D.}~\bibnamefont
  {Chandler}},\ }\href@noop {} {\bibfield  {journal} {\bibinfo  {journal} {J.
  Chem. Phys.}\ }\textbf {\bibinfo {volume} {141}},\ \bibinfo {pages} {18C519}
  (\bibinfo {year} {2014})}\BibitemShut {NoStop}%
\bibitem [{\citenamefont {Tarazona}\ and\ \citenamefont
  {Evans}(1982)}]{Tarazona:1982aa}%
  \BibitemOpen
  \bibfield  {author} {\bibinfo {author} {\bibfnamefont {P.}~\bibnamefont
  {Tarazona}}\ and\ \bibinfo {author} {\bibfnamefont {R.}~\bibnamefont
  {Evans}},\ }\href@noop {} {\bibfield  {journal} {\bibinfo  {journal} {Mol.
  Phys.}\ }\textbf {\bibinfo {volume} {47}},\ \bibinfo {pages} {1033} (\bibinfo
  {year} {1982})}\BibitemShut {NoStop}%
\bibitem [{\citenamefont {Evans}\ and\ \citenamefont
  {Parry}(1990)}]{Evans:1990aa}%
  \BibitemOpen
  \bibfield  {author} {\bibinfo {author} {\bibfnamefont {R.}~\bibnamefont
  {Evans}}\ and\ \bibinfo {author} {\bibfnamefont {A.~O.}\ \bibnamefont
  {Parry}},\ }\href@noop {} {\bibfield  {journal} {\bibinfo  {journal} {J.
  Phys.: Condens. Matter}\ }\textbf {\bibinfo {volume} {2}},\ \bibinfo {pages}
  {SA15} (\bibinfo {year} {1990})}\BibitemShut {NoStop}%
\bibitem [{\citenamefont {Gelfand}\ and\ \citenamefont
  {Fisher}(1990)}]{Gelfand:1990ve}%
  \BibitemOpen
  \bibfield  {author} {\bibinfo {author} {\bibfnamefont {M.~P.}\ \bibnamefont
  {Gelfand}}\ and\ \bibinfo {author} {\bibfnamefont {M.~E.}\ \bibnamefont
  {Fisher}},\ }\href {\doibase http://dx.doi.org/10.1016/0378-4371(90)90099-E}
  {\bibfield  {journal} {\bibinfo  {journal} {Physica A}\ }\textbf {\bibinfo
  {volume} {166}},\ \bibinfo {pages} {1 } (\bibinfo {year} {1990})}\BibitemShut
  {NoStop}%
\bibitem [{\citenamefont {Godawat}\ \emph {et~al.}()\citenamefont {Godawat},
  \citenamefont {Jamadagni}, \citenamefont {Venkateshwara},\ and\ \citenamefont
  {Garde}}]{Godawat:aa}%
  \BibitemOpen
  \bibfield  {author} {\bibinfo {author} {\bibfnamefont {R.}~\bibnamefont
  {Godawat}}, \bibinfo {author} {\bibfnamefont {S.}~\bibnamefont {Jamadagni}},
  \bibinfo {author} {\bibfnamefont {V.}~\bibnamefont {Venkateshwara}}, \ and\
  \bibinfo {author} {\bibfnamefont {S.}~\bibnamefont {Garde}},\ }\href@noop {}
  {}\bibinfo {note} {ArXiv:1409.2570}\BibitemShut {NoStop}%
\bibitem{thickfoot}
  \BibitemOpen
  \bibinfo {note} { One should also note that, in simulation, the maximum thickness of the drying (vapour) layer that can be investigated is only a few atomic diameters}\BibitemShut {NoStop}%
\bibitem [{\citenamefont {Parry}\ \emph
  {et~al.}(2008{\natexlab{a}})\citenamefont {Parry}, \citenamefont {Rasc\'on},
  \citenamefont {Bernardino},\ and\ \citenamefont
  {Romero-Enrique}}]{Parry:2008ef}%
  \BibitemOpen
  \bibfield  {author} {\bibinfo {author} {\bibfnamefont {A.~O.}\ \bibnamefont
  {Parry}}, \bibinfo {author} {\bibfnamefont {C.}~\bibnamefont {Rasc\'on}},
  \bibinfo {author} {\bibfnamefont {N.~R.}\ \bibnamefont {Bernardino}}, \ and\
  \bibinfo {author} {\bibfnamefont {J.~M.}\ \bibnamefont {Romero-Enrique}},\
  }\href {\doibase 10.1103/PhysRevLett.100.136105} {\bibfield  {journal}
  {\bibinfo  {journal} {Phys. Rev. Lett.}\ }\textbf {\bibinfo {volume} {100}},\
  \bibinfo {pages} {136105} (\bibinfo {year} {2008}{\natexlab{a}})}\BibitemShut
  {NoStop}%
\bibitem [{\citenamefont {Parry}\ \emph
  {et~al.}(2008{\natexlab{b}})\citenamefont {Parry}, \citenamefont
  {Rasc{\'o}n}, \citenamefont {Bernardino},\ and\ \citenamefont
  {Romero-Enrique}}]{Parry:2008rp}%
  \BibitemOpen
  \bibfield  {author} {\bibinfo {author} {\bibfnamefont {A.~O.}\ \bibnamefont
  {Parry}}, \bibinfo {author} {\bibfnamefont {C.}~\bibnamefont {Rasc{\'o}n}},
  \bibinfo {author} {\bibfnamefont {N.~R.}\ \bibnamefont {Bernardino}}, \ and\
  \bibinfo {author} {\bibfnamefont {J.~M.}\ \bibnamefont {Romero-Enrique}},\
  }\href {http://stacks.iop.org/0953-8984/20/i=49/a=494234} {\bibfield
  {journal} {\bibinfo  {journal} {J. Phys.: Condens. Matter}\ }\textbf
  {\bibinfo {volume} {20}},\ \bibinfo {pages} {494234} (\bibinfo {year}
  {2008}{\natexlab{b}})}\BibitemShut {NoStop}%
\bibitem [{\citenamefont {Parry}\ and\ \citenamefont
  {Rasc{\'o}n}(2009)}]{Parry:2009aa}%
  \BibitemOpen
  \bibfield  {author} {\bibinfo {author} {\bibfnamefont {A.~O.}\ \bibnamefont
  {Parry}}\ and\ \bibinfo {author} {\bibfnamefont {C.}~\bibnamefont
  {Rasc{\'o}n}},\ }\href {\doibase 10.1007/s10909-009-9902-2} {\bibfield
  {journal} {\bibinfo  {journal} {J. Low Temp. Phys.}\ }\textbf {\bibinfo
  {volume} {157}},\ \bibinfo {pages} {149} (\bibinfo {year}
  {2009})}\BibitemShut {NoStop}%
\bibitem [{\citenamefont {Evans}\ \emph {et~al.}(2016)\citenamefont {Evans},
  \citenamefont {Stewart},\ and\ \citenamefont {Wilding}}]{Evans:ab}%
  \BibitemOpen
  \bibfield  {author} {\bibinfo {author} {\bibfnamefont {R.}~\bibnamefont
  {Evans}}, \bibinfo {author} {\bibfnamefont {M.~C.}\ \bibnamefont {Stewart}},
  \ and\ \bibinfo {author} {\bibfnamefont {N.~B.}\ \bibnamefont {Wilding}},\
  }\href@noop {} {}\bibinfo {howpublished} {unpublished results} (\bibinfo
  {year} {2016})\BibitemShut {NoStop}%
\bibitem [{Note1()}]{Note1}%
  \BibitemOpen
  \bibinfo {note} {In a semi-infinite system, there is only a single
  vapor-liquid interface, which can wander arbitrarily far from the wall and
  the evaporation transition does not exist}\BibitemShut {NoStop}%
\bibitem [{\citenamefont {Mezger}\ \emph {et~al.}(2011)\citenamefont {Mezger},
  \citenamefont {Reichert}, \citenamefont {Ocko}, \citenamefont {Daillant},\
  and\ \citenamefont {Dosch}}]{Mezger:2011aa}%
  \BibitemOpen
  \bibfield  {author} {\bibinfo {author} {\bibfnamefont {M.}~\bibnamefont
  {Mezger}}, \bibinfo {author} {\bibfnamefont {H.}~\bibnamefont {Reichert}},
  \bibinfo {author} {\bibfnamefont {B.~M.}\ \bibnamefont {Ocko}}, \bibinfo
  {author} {\bibfnamefont {J.}~\bibnamefont {Daillant}}, \ and\ \bibinfo
  {author} {\bibfnamefont {H.}~\bibnamefont {Dosch}},\ }\href {\doibase
  10.1103/PhysRevLett.107.249801} {\bibfield  {journal} {\bibinfo  {journal}
  {Phys. Rev. Lett.}\ }\textbf {\bibinfo {volume} {107}},\ \bibinfo {pages}
  {249801} (\bibinfo {year} {2011})}\BibitemShut {NoStop}%
\bibitem [{\citenamefont {Uysal}\ \emph {et~al.}(2013)\citenamefont {Uysal},
  \citenamefont {Chu}, \citenamefont {Stripe}, \citenamefont {Timalsina},
  \citenamefont {Chattopadhyay}, \citenamefont {Schlep\"utz}, \citenamefont
  {Marks},\ and\ \citenamefont {Dutta}}]{Uysal:2013aa}%
  \BibitemOpen
  \bibfield  {author} {\bibinfo {author} {\bibfnamefont {A.}~\bibnamefont
  {Uysal}}, \bibinfo {author} {\bibfnamefont {M.}~\bibnamefont {Chu}}, \bibinfo
  {author} {\bibfnamefont {B.}~\bibnamefont {Stripe}}, \bibinfo {author}
  {\bibfnamefont {A.}~\bibnamefont {Timalsina}}, \bibinfo {author}
  {\bibfnamefont {S.}~\bibnamefont {Chattopadhyay}}, \bibinfo {author}
  {\bibfnamefont {C.~M.}\ \bibnamefont {Schlep\"utz}}, \bibinfo {author}
  {\bibfnamefont {T.~J.}\ \bibnamefont {Marks}}, \ and\ \bibinfo {author}
  {\bibfnamefont {P.}~\bibnamefont {Dutta}},\ }\href {\doibase
  10.1103/PhysRevB.88.035431} {\bibfield  {journal} {\bibinfo  {journal} {Phys.
  Rev. B}\ }\textbf {\bibinfo {volume} {88}},\ \bibinfo {pages} {035431}
  (\bibinfo {year} {2013})}\BibitemShut {NoStop}%
\bibitem [{\citenamefont {Nyg\aa{}rd}\ \emph {et~al.}(2016)\citenamefont
  {Nyg\aa{}rd}, \citenamefont {Sarman}, \citenamefont {Hyltegren},
  \citenamefont {Chodankar}, \citenamefont {Perret}, \citenamefont
  {Buitenhuis}, \citenamefont {van~der Veen},\ and\ \citenamefont
  {Kjellander}}]{Nygaard:2016aa}%
  \BibitemOpen
  \bibfield  {author} {\bibinfo {author} {\bibfnamefont {K.}~\bibnamefont
  {Nyg\aa{}rd}}, \bibinfo {author} {\bibfnamefont {S.}~\bibnamefont {Sarman}},
  \bibinfo {author} {\bibfnamefont {K.}~\bibnamefont {Hyltegren}}, \bibinfo
  {author} {\bibfnamefont {S.}~\bibnamefont {Chodankar}}, \bibinfo {author}
  {\bibfnamefont {E.}~\bibnamefont {Perret}}, \bibinfo {author} {\bibfnamefont
  {J.}~\bibnamefont {Buitenhuis}}, \bibinfo {author} {\bibfnamefont {J.~F.}\
  \bibnamefont {van~der Veen}}, \ and\ \bibinfo {author} {\bibfnamefont
  {R.}~\bibnamefont {Kjellander}},\ }\href {\doibase 10.1103/PhysRevX.6.011014}
  {\bibfield  {journal} {\bibinfo  {journal} {Phys. Rev. X}\ }\textbf {\bibinfo
  {volume} {6}},\ \bibinfo {pages} {011014} (\bibinfo {year}
  {2016})}\BibitemShut {NoStop}%
\bibitem [{\citenamefont {Nightingale}\ \emph {et~al.}(1983)\citenamefont
  {Nightingale}, \citenamefont {Saam},\ and\ \citenamefont
  {Schick}}]{Nightingale:1983ty}%
  \BibitemOpen
  \bibfield  {author} {\bibinfo {author} {\bibfnamefont {M.~P.}\ \bibnamefont
  {Nightingale}}, \bibinfo {author} {\bibfnamefont {W.~F.}\ \bibnamefont
  {Saam}}, \ and\ \bibinfo {author} {\bibfnamefont {M.}~\bibnamefont
  {Schick}},\ }\href {\doibase 10.1103/PhysRevLett.51.1275} {\bibfield
  {journal} {\bibinfo  {journal} {Phys. Rev. Lett.}\ }\textbf {\bibinfo
  {volume} {51}},\ \bibinfo {pages} {1275} (\bibinfo {year}
  {1983})}\BibitemShut {NoStop}%
\bibitem [{\citenamefont {Br\'ezin}\ \emph {et~al.}(1983)\citenamefont
  {Br\'ezin}, \citenamefont {Halperin},\ and\ \citenamefont
  {Leibler}}]{Brezin:1983dn}%
  \BibitemOpen
  \bibfield  {author} {\bibinfo {author} {\bibfnamefont {E.}~\bibnamefont
  {Br\'ezin}}, \bibinfo {author} {\bibfnamefont {B.~I.}\ \bibnamefont
  {Halperin}}, \ and\ \bibinfo {author} {\bibfnamefont {S.}~\bibnamefont
  {Leibler}},\ }\href {\doibase 10.1103/PhysRevLett.50.1387} {\bibfield
  {journal} {\bibinfo  {journal} {Phys. Rev. Lett.}\ }\textbf {\bibinfo
  {volume} {50}},\ \bibinfo {pages} {1387} (\bibinfo {year}
  {1983})}\BibitemShut {NoStop}%
\end{thebibliography}
 \end{document}